\documentstyle[amssymb,preprint,aps,psfig]{revtex}

\draft

\begin{document}
\title{Competing phases in the extended ${\bf U}$-${\bf V}$-${\bf J}$ Hubbard model
near the van Hove fillings}
\author{A.P. Kampf$^a$ and A.A. Katanin$^{a,b}$}
\address{$^a$ Institut f\"ur Physik, Theoretische Physik III,\\
Elektronische Korrelationen und Magnetismus,\\
Universit\"at Augsburg, 86135 Augsburg, Germany\\
$^b$ Institute of Metal Physics, 620219 Ekaterinburg, Russia}
\maketitle

\begin{abstract}
The phase diagram of the two-dimensional extended one-band $U$-$V$-$J$
Hubbard model is considered within a mean-field approximation and two- and
many-patch renormalization group (RG) approaches near the van Hove band
fillings. At small $t'$ and $J>0$ mean-field and many-patch RG approaches give
similar results for the leading spin-density-wave (SDW) instability,
while the two-patch RG approach, which predicts a wide region of charge-flux
(CF) phase becomes unreliable due to nesting effect. At the same
time, there is a complex competition between SDW, CF phases, and $d$-wave
superconductivity in two- and many-patch RG approaches. While the spin-flux
(SF) phase is not stable at the mean-field level, it is identified as a
possible ground state at $J<0$ in both RG approaches. With increasing $t'$
the results of all three approaches merge: d-wave superconductivity at $J>0$
and ferromagnetism at $J<0$ become the leading instabilities. For large enough
$V$ the charge-density-wave (CDW) state occurs. \end{abstract}

\vspace{0.2 in}


\section{Introduction}

Despite many years of research the mysterious properties of underdoped
high-temperature superconductors remain an unsolved and most challenging
problem of strongly correlated electron physics. Recent theoretical
proposals as well as experimental data trace the possible origin of the
unconventional electronic properties of underdoped cuprates to a delicate
competition between various types of ordering phenomena. Since metallic
cuprates result from doping Mott-Hubbard antiferromagnetic insulators, it is
natural that strong magnetic correlations persist for superconducting
compositions but alternative ordering tendencies may be hidden as well.

The relation between superconductivity and antiferromagnetism and the
attempt to treat both types of order on the same footing was the basis of
the phenomenological SO(5) theory as originally proposed by Zhang \cite{SO5}%
. Yet, the observation of inhomogeneous charge and spin structures in static
stripe patterns in rare-earth doped La$_{2-x}$Sr$_x$CuO$_4$ \cite{Stripes}
and the appearance of an anisotropic $d$-wave like pseudogap at temperatures
well above the superconducting transition have pointed to the possibility of
nearby instabilities towards other phases in competition or coexistence with
$d$-wave superconductivity.

The relevance of stripe patterns was proposed early on (for a review see,
e.g., Ref. \cite{Zaanen}), and years later their existence was indeed
experimentally verified in La$_{2-x-y}$Nd$_y$Sr$_x$CuO$_4$ \cite{Stripes}.
These static stripe structures may in fact result from the lattice
anisotropy connected to the CuO$_6$ octahedral tilt pattern in the LTT phase
of this compound\cite{Stripe1}. Time-reversal symmetry breaking charge-flux
(CF) \cite{Rice,OAF,AfflMarst} and spin-flux (SF, spin nematic) \cite
{Neresjan,SF,SF1} phases were also proposed early on as possible ground
states of the basic correlated electron Hamiltonians commonly used to model
high-T$_c$ superconductors.

The idea of circulating orbital currents (also called an orbital
antiferromagnet state\cite{Rice}) was recently revived as the underlying
origin of the pseudogap phase in underdoped cuprates\cite{DDW1}. Two
variants of orbital current structures are currently discussed: Varma \cite
{Varma} proposed a circulating current pattern in the CuO$_2$ planes which
preserves the discrete translational symmetry of the lattice, while the more
recently proposed $d$-density wave (DDW) state \cite{DDW1,DDW} contains a
staggered current pattern which doubles the unit cell. The latter state is
in fact equivalent to the orbital antiferromagnet discussed a long time ago%
\cite{OAF}. These time reversal symmetry breaking phases may furthermore
coexist with $d$-wave superconductivity. Yet, another instability connected
to the spontaneous deformation of the Fermi surface (Pomeranchuk
instability, PI) was recently proposed for the two-dimensional Hubbard model%
\cite{Metz}. It appears tempting to suggest a possible relation between the
Pomeranchuk instability and stripe pattern formation.

Remarkably, the possible existence of orbital currents has most recently
received experimental support. Angular-resolved photoemission with
circularly polarized light identified intensity differences for left and
right circularly polarized photons below the pseudogap temperature in Bi2212
\cite{Kaminski} and c-axis oriented ordered magnetic moments were detected
by spin-polarized neutron scattering in YBCO \cite{Mook}. Both data sets can
find a natural interpretation in terms of planar circulating current phases.

The tendency towards the formation of charge flux phases is most likely
beyond the physics of the standard Hubbard model. In order to allow for a
larger variety of competing or coexisting phases it appears therefore
demanding to explore some extensions of the Hubbard model. A natural choice
is to consider the extended $U$-$V$-$J$ Hubbard model ($V$ is a
nearest-neighbor interaction and $J$ is the Heisenberg exchange coupling).
This model was chosen early on to study orbital antiferromagnetism and spin
nematic phases. In one dimension (1D), this model was studied within the
bosonization technique \cite{UVJ1D,UVJ1D1}. However, in 1D, one can not
explore the possibility of charge- and spin-flux phases, but they were later
on investigated in an extension to ladder systems\cite{Ladder}. In two
dimensions, the $U$-$V$-$J$ model was considered within mean-field
approximation for $J=0$ \cite{MFV,Murakami} and for $V=0$ \cite{MFJ}. The
possible coexistence of antiferromagnetism and $d$-wave superconductivity%
\cite{AFSC}, and charge flux and $d$-wave superconductivity \cite{CFSC} was
discussed in this context. More recently, the possibility of coexisting
phases was reanalyzed within a mean-field treatment of the extended Hubbard
model which furthermore includes correlated hopping \cite{CFSC1}.

Common to these studies is the need to go beyond the standard one-band
Hubbard model to allow for a variety of alternative ordering phenomena.
However, a mean-field treatment does not allow for an accurate treatment of
fluctuations which may alter the complex ground-state phase diagram. Other
analytical and numerical methods are needed for a reliable insight into the
interplay of various types of ordering tendencies. One possible step forward
in this direction has been the recent analysis of competing phases within
dynamical mean-field theory \cite{AFSC-DMFT}.

Another possible route is to start from the weak-coupling regime, where
magnetic or superconducting instabilities can result from nesting of the
Fermi surface (FS) or van Hove singularities (vHS) in the density of states.
The latter case is also relevant from a material oriented point of view
since the FS of cuprates in an intermediate doping regime is close to vHS of
the electron spectrum\cite{ARPES,ARPES1,ARPES2}. Schulz considered \cite
{Schulz} different phases near van Hove band fillings within a
mean-field-like analysis which, however, did not take into account the
interplay of different electron scattering channels. The parquet approach
was first applied to the extended Hubbard model with a nearest-neighbor
hopping in Refs. \cite{Dzy1,DzyYak,DzyYak1}, where besides the standard
superconducting, spin- and charge-density-wave instabilities, the
possibilities for spin- and charge-flux instabilities were also discussed.
Recently, the standard Hubbard model ($V=J=0$) was investigated within the
renormalization-group (RG) approach on a patched FS using the Polchinsky
flow equations \cite{ZanSchulz}, Wick-ordered RG equations \cite{Metzner},
and a RG scheme for one-particle irreducible functions \cite{SalmHon}. The
possibility of the charge-flux phase as well as the Pomeranchuk instability
for the standard Hubbard model was addressed in Ref. \cite{Hon2002}. These
latter methods have the advantage that they allow to take into account the
contributions of the entire FS. At the same time, even the restriction of
the momenta to the vicinity of van Hove points (the so called ``two-patch''
approach originally proposed in Refs. \cite{Led,FurRice}) can capture
already the essential physics and can give results which are in qualitative
agreement with the RG approaches on patched Fermi surfaces, see Ref. \cite
{OurVH}. This two-patch approach allows to include the contribution of
particle-hole scattering at small momenta which is expected to be important
for the determination of the complete ground-state phase diagram. The role
of this type of scattering was also explored within the RG approach with a
temperature cutoff on the patched FS \cite{SalmHon1}. Although the phase
diagram of the $U$-$V$-$J$ Hubbard model was investigated previously within
the two-patch RG approach, only the cases of $J=0$ \cite{Murakami} and
arbitrary $J$ with nearest-neighbor hopping dispersion\cite{Binz}, but
without the contribution of the small-momenta particle-hole scattering were
considered. The same model within the many-patch approach was not
systematically studied so far.

In this paper we reconsider the phase diagram of the extended $U$-$V$-$J$
Hubbard model with special emphasis on the possibility of flux phases$.$ We
employ the mean-field approximation, two- and many-patch RG approaches. The
paper is organized as follows: In Sect. II we start with the mean-field
phase diagram of the $U$-$V$-$J$ Hubbard model. In Sect. III we describe the
two- and many-patch RG schemes and compare the results obtained with both
approaches to the mean-field phase diagram. In Sect. IV we complement the RG
analysis by symmetry arguments. Finally, in Sect. V we summarize the main
results and conclude.

\section{Mean-field analysis}

We consider the extended $U$-$V$-$J$ Hubbard model on a square lattice as
given by
\begin{equation}
H=-\sum_{ij\sigma }t_{ij}c_{i\sigma }^{\dagger }c_{j\sigma
}+U\sum_in_{i\uparrow }n_{i\downarrow }+\frac V2\sum_{\langle ij\rangle
}n_in_j+\frac J2\sum_{\langle ij\rangle }{\bf S}_i\cdot {\bf S}_j-\mu N_e
\label{H}
\end{equation}
where the hopping amplitude $t_{ij}=t$ for nearest neighbor sites $i$ and $j$
and $t_{ij}=-t^{\prime }$ for next-nearest neighbor sites ($t,t^{\prime }>0$%
); $N_e$ is the total number of electrons. $U>0$ is the on-site and $V$ the
nearest-neighbor interaction, $J$ is the Heisenberg exchange coupling, and
\[
n_i=\sum_\sigma c_{i\sigma }^{\dagger }c_{i\sigma },\;{\bf S}_i=\frac 12%
\sum_{\sigma \sigma ^{\prime }}c_{i\sigma }^{\dagger }%
\mbox {\boldmath
$\sigma $}_{\sigma \sigma ^{\prime }}c_{i\sigma ^{\prime }};
\]
$\mbox {\boldmath $\sigma $}_{\sigma \sigma ^{\prime }}$ are the Pauli
matrices. We introduce operators which correspond to different types of
order
\begin{eqnarray}
\widehat{O}_{{\rm CDW}}(i) &=&\frac 12(-1)^i\sum_\sigma (c_{i\sigma
}^{\dagger }c_{i\sigma }-n/2),  \nonumber \\
\widehat{O}_{{\rm SDW}}(i) &=&\frac 12(-1)^i\sum_\sigma \sigma (c_{i\sigma
}^{\dagger }c_{i\sigma }-n/2),  \nonumber \\
\widehat{O}_{{\rm dSC}}(i) &=&\frac 1{2z}\sum_{j\sigma }\lambda _{ij}\sigma
c_{i\sigma }^{\dagger }c_{j,-\sigma }^{\dagger },  \nonumber \\
\widehat{O}_{{\rm CF}}(i) &=&-\frac{\text{i}}{2z}(-1)^i\sum_{j\sigma
}\lambda _{ij}c_{i\sigma }^{\dagger }c_{j\sigma },  \nonumber \\
\widehat{O}_{{\rm SF}}(i) &=&-\frac{\text{i}}{2z}(-1)^i\sum_{j\sigma
}\lambda _{ij}\sigma c_{i\sigma }^{\dagger }c_{j\sigma },  \nonumber \\
\widehat{O}_{{\rm F}}(i) &=&\frac 12\sum_\sigma \sigma (c_{i\sigma
}^{\dagger }c_{i\sigma }-1/2),  \label{O}
\end{eqnarray}
where $\lambda _{ij}=+1\;(-1)$ for $j=i+\delta _x(\delta _y)$ and zero
otherwise; $\delta _{x,y}$ denote unit vectors which connect to nearest
neighbor sites in $x\ $and $y$ directions respectively. $(-1)^i\equiv
(-1)^{i_x+i_y},$ $z=4$ is the number of nearest neighbor sites on the square
lattice, and $n$ is the electron density. Nonzero average values of the
operators $\widehat{O}_{{\rm CDW}}(i)$ and $\widehat{O}_{{\rm SDW}}(i)$
correspond to charge- and spin density wave states, $\widehat{O}_{{\rm dSC}%
}(i)$ to $d$-wave superconductivity, $\widehat{O}_{{\rm CF}}$ and $\widehat{O%
}_{{\rm SF}}$ to states with charge flux (the orbital antiferromagnet state)
\cite{Rice,OAF,DDW} or spin flux (spin nematic state) \cite{Neresjan,SF,SF1}%
, respectively, and finally $\widehat{O}_{{\rm F}}$ corresponds to a
ferromagnetic state. We decouple the interactions by introducing the order
parameters $O_m=\langle \widehat{O}_m\rangle $ ($m={\rm CDW,SDW,}...$ ), and
arrive at the mean-field Hamiltonian
\begin{equation}
H=\sum_{{\bf k}\sigma }\varepsilon _{{\bf k}}c_{{\bf k}\sigma }^{\dagger }c_{%
{\bf k}\sigma }+\sum_{{\bf k}}\Delta _m\widehat{O}_m({\bf k})+\frac{\Delta
_m^2}{\Gamma _m}
\end{equation}
where
\begin{equation}
\varepsilon _{{\bf k}}=-2t(\cos k_x+\cos k_y)+4t^{\prime }\cos k_x\cos
k_y-\mu .  \label{ek}
\end{equation}
$\;\widehat{O}_m({\bf k})$ are the Fourier transforms of the operators in
Eq. (\ref{O}), $\Delta _m=O_m\Gamma _m$ and
\begin{equation}
\begin{array}{lll}
\Gamma _{{\rm CDW}}=8V-U; & \Gamma _{{\rm SDW}}=U+2J; & \Gamma _{{\rm dSC}%
}=3J-4V; \\
\Gamma _{{\rm CF}}=3J+4V; & \Gamma _{{\rm SF}}=4V-J; & \Gamma _{{\rm F}%
}=U-2J.
\end{array}
\label{GMF}
\end{equation}

At finite temperature $T$ the self-consistency conditions lead to the
mean-field equations (see e.g. Ref. \cite{Neresjan})
\begin{eqnarray}
1 &=&\frac{\Gamma _m}{2N}\sum_{{\bf k},\alpha =\pm }^{\prime }\frac{\phi _{%
{\bf k}}^2}{E_{{\bf k}\alpha }^i}\tanh \frac{E_{{\bf k}\alpha }^m}{2T},
\nonumber \\
n &=&1-\frac 1{2N}\sum_{{\bf k},\alpha =\pm }^{\prime }\tanh \frac{E_{{\bf k}%
\alpha }^m}{2T},  \label{MF1}
\end{eqnarray}
where $\Sigma ^{\prime }$ denotes the summation over momenta in the magnetic
Brillouin zone defined by $\cos k_x+\cos k_y\geq 0$ (the lattice constant
has been set to unity); $\phi _{{\bf k}}=1$ for SDW and CDW phases and $\phi
_{{\bf k}}=f_{{\bf k}}\equiv (\cos k_x-\cos k_y)/2$ for CF, SF, and dSC
phases,
\begin{equation}
E_{{\bf k}\alpha }^m=\frac{\varepsilon _{{\bf k}}+\varepsilon _{{\bf k+Q}}}2+%
\frac \alpha 2\sqrt{(\varepsilon _{{\bf k}}-\varepsilon _{{\bf k+Q}%
})^2+4\phi _{{\bf k}}^2\Delta _m^2}
\end{equation}
for $m={\rm SDW,CDW,CF}$ or {\rm SF} and
\begin{equation}
E_{{\bf k}\alpha }^{{\rm dSC}}=\frac 12\sqrt{(\varepsilon _{{\bf k}%
}-\varepsilon _{{\bf k+Q}})^2+4\phi _{{\bf k}}^2\Delta _{{\rm dSC}}^2}.
\end{equation}
${\bf Q}=(\pi ,\pi )\ $is the wavevector for staggered order. For the
ferromagnetic phase the mean-field equations read
\begin{eqnarray}
\Delta _{{\rm F}} &=&\frac{\Gamma _{{\rm F}}}N\sum_{{\bf k,}\alpha =\pm
}\alpha f_{{\bf k}\alpha },\;n=\frac 1N\sum_{{\bf k}\alpha }f_{{\bf k}\alpha
},  \nonumber \\
E_{{\bf k}\alpha }^{{\rm F}} &=&\varepsilon _{{\bf k}}-\alpha \Delta _{{\rm F%
}}  \label{MFF}
\end{eqnarray}
where $f_{{\bf k}\alpha }\equiv f(E_{{\bf k}\alpha }^{{\rm F}})$ is the
Fermi function. Non-trivial mean-field solutions exist only for $\Gamma
_m>0, $ which therefore determines the borders of absolute stability of the
corresponding phases in the phase diagram.

We also allow for coexistence of antiferromagnetism, $d$-wave
superconductivity, and charge-flux phases. The corresponding MF equations
are the generalization of those for the coexistence of SDW and dSC phases
\cite{AFSC} and CF and dSC phases \cite{CFSC}, and have the form
\begin{eqnarray}
1 &=&\frac{\Gamma _{{\rm SDW}}}{2N}\sum_{{\bf k},\alpha =\pm }^{\prime }%
\frac 1{E_{{\bf k}\alpha }}\tanh \frac{E_{{\bf k}\alpha }}{2T}\left( 1+\frac{%
\varepsilon _{{\bf k}}+\varepsilon _{{\bf k+Q}}}{g_{k\alpha }}\right) ,
\nonumber \\
1 &=&\frac{\Gamma _{{\rm dSC}}}{2N}\sum_{{\bf k},\alpha =\pm }^{\prime }%
\frac{f_{{\bf k}}^2}{E_{{\bf k}\alpha }}\tanh \frac{E_{{\bf k}\alpha }}{2T},
\nonumber \\
1 &=&\frac{\Gamma _{{\rm CF}}}{2N}\sum_{{\bf k},\alpha =\pm }^{\prime }\frac{%
f_{{\bf k}}^2}{E_{{\bf k}\alpha }}\tanh \frac{E_{{\bf k}\alpha }}{2T}\left(
1+\frac{\varepsilon _{{\bf k}}+\varepsilon _{{\bf k+Q}}}{g_{k\alpha }}%
\right) ,  \nonumber \\
n &=&1-\frac 1{2N}\sum_{{\bf k},\alpha =\pm }^{\prime }\frac 1{E_{{\bf k}%
\alpha }}\tanh \frac{E_{{\bf k}\alpha }}{2T}\left( \varepsilon _{{\bf k}%
}+\varepsilon _{{\bf k+Q}}+g_{k\alpha }\right) ,  \label{MF2}
\end{eqnarray}
where the electronic spectrum is given by
\begin{eqnarray}
E_{{\bf k}\alpha } &=&\sqrt{(\varepsilon _{{\bf k}}^2+\varepsilon _{{\bf k+Q}%
}^2)/2+\Delta _{{\rm SDW}}^2+f_{{\bf k}}^2(\Delta _{{\rm CF}}^2+\Delta _{%
{\rm dSC}}^2)+(\varepsilon _{{\bf k}}+\varepsilon _{{\bf k+Q}})g_{k\alpha }/2%
},  \nonumber \\
g_{{\bf k}\alpha } &=&\alpha \sqrt{(\varepsilon _{{\bf k}}-\varepsilon _{%
{\bf k+Q}})^2+4(\Delta _{{\rm SDW}}^2+f_{{\bf k}}^2\Delta _{{\rm dSC}}^2)}.
\end{eqnarray}
We note that a nonzero expectation value of both, antiferromagnetic and
superconducting order parameters, produces also a nonzero $\pi $-triplet
pairing amplitude with momentum ${\bf Q}$ (see e.g. Ref.\cite{AFSC}).
Analogously nonzero values of charge flux and $d$-wave superconducting order
parameters lead to a finite singlet pairing amplitude with momentum ${\bf Q}$
(so called $\eta $-pairing). Here, we do not include $\pi $- and $\eta $%
-pairing in the mean-field analysis, since it was verified numerically\cite
{AFSC} that the corresponding amplitudes are significantly smaller than the
amplitudes of the corresponding ``parent'' order parameters.

The mean-field equations (\ref{MF1}), (\ref{MFF}), and (\ref{MF2}) were
solved numerically for $T=0,$ $U=4t,$ $t^{\prime }=0.1t,$ and band filling $%
n=0.92$ (which is the van Hove filling for this value of $t^{\prime }/t,$
see the next section). Since the regions of stability of different phases
overlap, we compare the energies $E=\langle H\rangle +\mu N_e$ to identify
the mean-field ground state. The resulting phase diagram is shown in Fig. 1.
At positive $J$ and not too large $V$ we find coexistence of a spin density
wave with $d$-wave superconductivity. However, the fraction of
superconductivity in this phase is quite small (for $V=0,$ $J=0.5U$ we have $%
O_{{\rm SDW}}=0.40$ and $O_{{\rm dSC}}=0.06$). With decreasing $V$ the
superconducting fraction increases and becomes substantial close to the
transition into the pure dSC state (for $V=-3U/4,$ $J=-0.05U$ we have $O_{%
{\rm SDW}}=0.18$ and $O_{{\rm dSC}}=0.12$), and finally at the boundary
between the dSC and SDW+dSC phases the antiferromagnetic order parameter
vanishes continuously. Therefore, this transition as well as the transition
from the SDW into the SDW+dSC phase is of second order. All other phase
boundary lines in Fig. 1 correspond to first order transitions. The CF, SF,
and CF+$d$SC phases are not energetically favorable at all on the mean-field
level, the dSC phase is stable only for $V<-2t.$ The mean-field phase
diagram at other fillings or other $t^{\prime }/t$ is similar. In the
special case $t^{\prime }=0$ and half-filling, the coexistence region of the
SDW+dSC phase is replaced by a pure SDW state.

Away from half-filling the phases with ordering wavevector ${\bf Q=(}\pi
,\pi ),$ i.e. CDW, SDW, CF, and SF phases are usually unstable towards phase
separation (see e.g. the discussion in Refs. \cite{Dongen,Guinea}). To
analyze this possibility, we calculate the isothermal compressibility $%
\kappa _T$ defined by
\begin{equation}
\kappa _T=\frac 1{n^2}\frac{\partial n}{\partial \mu }=\frac N{n^2}\left(
\frac{\partial ^2E}{\partial n^2}\right) ^{-1}.
\end{equation}
A negative $\kappa _T$ necessarily implies the thermodynamic instability of
the considered homogeneous phase and a possible tendency towards phase
separation. The numerical analysis shows that all above phases are indeed
unstable except for the SDW+dSC phase in the region $J\ll U$ and $%
V\rightarrow -U$ where the fraction of the SDW order parameter is small.

The above mean-field results provide a rough qualitative picture for the
regions of different stable phases in the ground-state phase diagram of the $%
U$-$V$-$J$ Hubbard model. More detailed information is extracted from the
weak-coupling RG treatments discussed below.

\section{Renormalization-group analysis}

\subsection{Two-patch analysis}

The tight-binding spectrum (\ref{ek}) leads to vHS in the density of states
arising from the contributions of the points ${\bf k}_A=(\pi ,0)\ $and ${\bf %
k}_B=(0,\pi ).$ These singularities lie at the FS if $\mu =-4t^{\prime }$.
For $t^{\prime }=0$ the corresponding filling is $n_{VH}=1$ and the FS is
nested, but the nesting is removed for $t^{\prime }/t>0,$ when $n_{VH}<1.$
The shape of the FS at different $t^{\prime }/t$ and van Hove band fillings
is shown in Fig. 2.

The two-patch approach\cite{Led,FurRice,OurVH} considers the fillings which
are close to the van Hove band fillings. At these fillings the density of
states at the Fermi energy and the electron-electron interaction vertices at
momenta ${\bf k}={\bf k}_{A,B}$ contain logarithmical divergencies coming
from the momentum integrations in the vicinity of the points ${\bf k}={\bf k}%
_{A,B}$, and therefore one can expect that the contributions from the
vicinities of these points are the most important for the calculation of the
renormalized electron-electron interaction vertices.

We start by rewriting the $U$-$V$-$J$ Hamiltonian in momentum space
\begin{equation}
H=\sum_{{\bf k}\sigma }\varepsilon _{{\bf k}}c_{{\bf k}\sigma }^{\dagger }c_{%
{\bf k}\sigma }+\frac 1{2N^2}\sum_{{\bf k}_1{\bf k}_2{\bf k}_3{\bf k}%
_4}\sum_{\sigma \sigma ^{\prime }}g({\bf k}_1,{\bf k}_2,{\bf k}_3,{\bf k}%
_4)c_{{\bf k}_1\sigma }^{\dagger }c_{{\bf k}_2\sigma ^{\prime }}^{\dagger
}c_{{\bf k}_3\sigma ^{\prime }}c_{{\bf k}_4\sigma }\delta _{{\bf k}_1+{\bf k}%
_2-{\bf k}_3-{\bf k}_4}  \label{H1}
\end{equation}
where
\begin{equation}
g({\bf k}_1,{\bf k}_2,{\bf k}_3,{\bf k}_4)=U+(V-J/4)\gamma _{{\bf k}_2-{\bf k%
}_3}-(J/2)\gamma _{{\bf k}_3-{\bf k}_1}
\end{equation}
$\gamma _{{\bf k}}=2(\cos k_x+\cos k_y)$ and the Kronecker $\delta $-symbol
ensures momentum conservation$.$ Since we restrict the momenta to the
vicinity of ${\bf k}_A$ and ${\bf k}_B$, it is convenient to introduce new
electron operators $a_{{\bf k}}$ and $b_{{\bf k}}$ by
\[
c_{{\bf k}\sigma }=\left\{
\begin{array}{cc}
a_{{\bf k-k}_A,\sigma } & {\bf k}\in O(A) \\
b_{{\bf k-k}_B,\sigma } & {\bf k}\in O(B)
\end{array}
\right.
\]
here $O(A)=\{{\bf k}:\,|{\bf k}-{\bf k}_A|<\Lambda \}$ and similar for $B;$ $%
\Lambda $ is a momentum cutoff parameter. We expand the spectrum near the
van Hove points
\begin{mathletters}
\begin{eqnarray}
\varepsilon _{{\bf k}_A+{\bf p}} &\equiv &\varepsilon _{{\bf p}}^A=-2t(\sin
^2\varphi \,p_x^2-\cos ^2\varphi \,p_y^2)-\widetilde{\mu }  \label{eka} \\
\ &=&-2tp_{+}p_{-}-\widetilde{\mu }  \nonumber \\
\varepsilon _{{\bf k}_B+{\bf p}} &\equiv &\varepsilon _{{\bf p}}^B=2t(\cos
^2\varphi \,p_x^2-\sin ^2\varphi \,p_y^2)-\widetilde{\mu }  \label{ekb} \\
\ &=&2t\widetilde{p}_{+}\widetilde{p}_{-}-\widetilde{\mu }  \nonumber
\end{eqnarray}
where $\cos (2\varphi )=R=2t^{\prime }/t,\ \widetilde{\mu }=\mu +4t^{\prime
},$ $p_{\pm }=p_x\sin \varphi \pm p_y\cos \varphi ,$ $\widetilde{p}_{\pm
}=p_x\sin \varphi \pm p_y\cos \varphi $ and rewrite the Hamiltonian in the
form
\end{mathletters}
\begin{eqnarray}
H &=&\sum_{{\bf p}\sigma }\varepsilon _{{\bf p}}^Aa_{{\bf p}\sigma
}^{\dagger }a_{{\bf p}\sigma }+\sum_{{\bf p}\sigma }\varepsilon _{{\bf p}%
}^Bb_{{\bf p}\sigma }^{\dagger }b_{{\bf p}\sigma }  \nonumber \\
&&\ \ \ +\frac{2\pi ^2t}{N^2}\sum_{{\bf p}_i,\sigma \sigma ^{\prime
}}[g_1(\lambda )a_{{\bf p}_1\sigma }^{\dagger }b_{{\bf p}_2\sigma ^{\prime
}}^{\dagger }a_{{\bf p}_3\sigma ^{\prime }}b_{{\bf p}_4\sigma }+g_2(\lambda
)a_{{\bf p}_1\sigma }^{\dagger }b_{{\bf p}_2\sigma ^{\prime }}^{\dagger }b_{%
{\bf p}_3\sigma ^{\prime }}a_{{\bf p}_4\sigma }]\delta _{{\bf p}_1+{\bf p}_2-%
{\bf p}_3-{\bf p}_4}  \nonumber \\
&&\ \ \ +\frac{\pi ^2t}{N^2}\sum_{{\bf p}_i,\sigma \sigma ^{\prime
}}[g_3(\lambda )a_{{\bf p}_1\sigma }^{\dagger }a_{{\bf p}_2\sigma ^{\prime
}}^{\dagger }b_{{\bf p}_3\sigma ^{\prime }}b_{{\bf p}_4\sigma }+g_4(\lambda
)a_{{\bf p}_1\sigma }^{\dagger }a_{{\bf p}_2\sigma ^{\prime }}^{\dagger }a_{%
{\bf p}_3\sigma ^{\prime }}a_{{\bf p}_4\sigma }+a
\begin{array}{c}
\leftrightarrow
\end{array}
b]\delta _{{\bf p}_1+{\bf p}_2-{\bf p}_3-{\bf p}_4}  \label{HVH}
\end{eqnarray}
where $\varepsilon _{{\bf p}}^{A,B}=\varepsilon _{{\bf k}_{A,B}+{\bf p}},$ $%
\lambda =\ln (\Lambda /\max (p_{i+},p_{i-},\widetilde{p}_{i+},\widetilde{p}%
_{i-},|\mu |/t);$ the summation is restricted to momenta $|{\bf p}%
_i|<\Lambda .$ Neglecting the weak (non-logarithmical) dependence of the
vertices $g_i$ on momenta at $|{\bf p}_i|<\Lambda $, we obtain the bare
values of the vertices $g_i^0$ which are independent of $\lambda $ and given
by
\begin{eqnarray}
g_1^0 &=&g({\bf k}_A,{\bf k}_B,{\bf k}_A,{\bf k}_B)=g_0(1-4V/U-J/U),
\nonumber \\
g_2^0 &=&g({\bf k}_A,{\bf k}_B,{\bf k}_B,{\bf k}_A)=g_0(1+4V/U+J/U),
\nonumber \\
g_3^0 &=&g({\bf k}_A,{\bf k}_A,{\bf k}_B,{\bf k}_B)=g_0(1-4V/U+3J/U),
\nonumber \\
g_4^0 &=&g({\bf k}_A,{\bf k}_A,{\bf k}_A,{\bf k}_A)=g_0(1+4V/U-3J/U).
\end{eqnarray}
where $g_0=U/(4\pi ^2t)$ is the dimensionless coupling parameter. Note that
at half filling and $t^{\prime }=0$ the contributions of the flat, nested
parts of the FS (which should be considered separately, see, e.g. Ref. \cite
{DzyYak1} and references therein) is neglected. The contribution of these
nested parts of the Fermi surface can be considered within the many-patch
approach, as we discuss in the next section.

To obtain the dependence of the vertices $g_i$ on $\lambda $ we apply the RG
analysis. We start with the bare values of the vertices at momenta far from
van Hove singularities, i.e. with $g_i^0,$ and integrate out the fermions $%
a_{{\bf p}}$ with momenta $\Lambda e^{-\lambda }<p_{\pm }<\Lambda
e^{-\lambda -d\lambda }$ and fermions $b_{{\bf p}}$ with momenta $\Lambda
e^{-\lambda }<\widetilde{p}_{\pm }<\Lambda e^{-\lambda -d\lambda }$ at each
RG step. We consider first the one-loop corrections which contain
particle-hole ($ph$) and particle-particle ($pp$) bubbles at small momenta
and momenta close to ${\bf Q}.$ The results for these bubbles can be
summarized as follows:
\begin{mathletters}
\label{Hi0}
\begin{eqnarray}
\Pi _{{\bf q}}^{pp} &=&\sum_{\Lambda e^{-\lambda }<p_{\pm }<\Lambda }\frac{%
1-f(\varepsilon _{{\bf p}}^A)-f(\varepsilon _{{\bf p+q}}^A)}{\varepsilon _{%
{\bf p}}^A+\varepsilon _{{\bf p+q}}^A}=\frac{c_0}{4\pi ^2t}\lambda ^2,
\label{Hi0c} \\
\Pi _{{\bf q+Q}}^{ph} &=&\sum_{\Lambda e^{-\lambda }<p_{\pm }<\Lambda }\frac{%
f(\varepsilon _{{\bf p}}^A)-f(\varepsilon _{{\bf p+q}}^B)}{\varepsilon _{%
{\bf p}}^A-\varepsilon _{{\bf p+q}}^B}=\frac 1{4\pi ^2t}\min (\lambda ^2,2z_{%
{\bf Q}}), \\
\Pi _{{\bf q+Q}}^{pp} &=&\sum_{\Lambda e^{-\lambda }<p_{\pm }<\Lambda }\frac{%
1-f(\varepsilon _{{\bf p}}^A)-f(\varepsilon _{{\bf p+q}}^B)}{\varepsilon _{%
{\bf p}}^A+\varepsilon _{{\bf p+q}}^B}=\frac{c_{{\bf Q}}}{2\pi ^2t}\lambda ,
\end{eqnarray}
where
\end{mathletters}
\begin{eqnarray}
c_0 &=&1/\sin (2\varphi )=1/\sqrt{1-R^2},  \nonumber \\
c_{{\bf Q}} &=&\tan ^{-1}(R/\sqrt{1-R^2})/R,  \nonumber \\
z_{{\bf Q}} &=&\ln [(1+\sqrt{1-R^2})/R].
\end{eqnarray}
The contribution of the slices $\Lambda e^{-\lambda }<p_{\pm },\widetilde{p}%
_{\pm }<\Lambda e^{-\lambda -d\lambda }$ is obtained by taking the
derivatives $d\Pi _i/d\lambda $. Since the contribution to the particle-hole
($\Pi _{{\bf q}}^{ph}$) channel is concentrated near the FS, we use the
weaker cutoff condition $\Lambda e^{-\lambda }<p_{+}<\Lambda $ or $\Lambda
e^{-\lambda }<p_{-}<\Lambda $ for this channel which gives
\begin{equation}
\Pi _{{\bf q}}^{ph}=\sum_{\Lambda e^{-\lambda }<p_{+}<\Lambda \text{ or }%
\Lambda e^{-\lambda }<p_{-}<\Lambda }\frac{f(\varepsilon _{{\bf p}%
}^A)-f(\varepsilon _{{\bf p+q}}^A)}{\varepsilon _{{\bf p}}^A-\varepsilon _{%
{\bf p+q}}^A}=\frac{z_0\lambda }{2\pi ^2t}  \label{Pph}
\end{equation}
with $z_0=c_0.$ A more accurate treatment of the $ph$ channel requires the
parquet summation of the one-loop diagrams, it was shown, however, in Ref.
\cite{OurVH} that the corresponding corrections do not change qualitatively
the results. Equation (\ref{Pph}) can be justified also by considering the
temperature flow of the vertices on a patched FS as proposed in Ref.\cite
{SalmHon1} (see also the next section).

While the bubbles $\Pi _{{\bf Q}}^{ph}$ (at $\lambda <2z_{{\bf Q}}$) and $%
\Pi _0^{pp}$ contain squared logarithms, the bubbles $\Pi _{{\bf Q}}^{ph}$
(at $\lambda >2z_{{\bf Q}}$)$,$ $\Pi _0^{ph}$ and $\Pi _{{\bf Q}}^{pp}$
contain only single-logarithmical divergencies. Strictly speaking, the RG
approach does not perform a correct summation of these subleading
divergencies. However, one can not simply neglect the single-logarithmical
contributions ($\Pi _{{\bf q+Q}}^{pp}$ at $\lambda <2z_{{\bf Q}},$ $\Pi
_0^{ph}$ and $\Pi _{{\bf q}}^{ph}$) in comparison to the
squared-logarithmical ones, since it amounts to ignore some electron
scattering channels and therefore would exclude the possibility of related
orders, see Ref. \cite{OurVH} for a discussion. Mathematically, this
difficulty manifests itself in the corresponding phases by the growing of
the vertices which are multiplied by single logarithms, whose contribution
can therefore become comparable to the squared-logarithmic contributions\cite
{OurVH}.

As we will see below from the solution of the RG equations, in the
weak-coupling regime at small{\it \ }$t^{\prime }/t\ll 1$ the contribution
of the subleading corrections $\Pi _0^{ph}$ and $\Pi _{{\bf Q}}^{pp}$ is
small so that they do not change substantially the flow of the coupling
constants. With increasing coupling, the contribution of
single-logarithmical terms is comparable to the contribution of the leading
squared-logarithmical terms which reflects the possibility of a
corresponding ordering phenomena (e.g. ferromagnetism) in the
strong-coupling regime, although the latter can not be explored within the
weak-coupling RG approach.

At intermediate $t^{\prime }/t$ the situation changes. In this case the
contributions of different channels (containing single- and
double-logarithmical terms) are comparable already in the weak-coupling
regime. We have checked however, that even for intermediate $t^{\prime }/t$
the qualitative results of the two-patch RG approach do not depend on the
cutoff procedure, which means that the treatment of the subleading
single-logarithmical terms within the same RG procedure for intermediate $%
t^{\prime }/t$ is qualitatively reliable. Therefore, we take into account
the contribution of both, single- and squared-logarithmical terms in Eqs. (%
\ref{Hi0}) and (\ref{Pph}). At the same time, the contribution of the
subleading single-logarithmical term in Eq. (\ref{Hi0c}) can be safely
neglected, since it is multiplied by the same vertex as the leading one and
therefore is always subleading.

We determine the RG equations for the vertices $g_i(\lambda )$ in the form
\cite{Led,FurRice,OurVH}
\begin{eqnarray}
{\rm d}g_1/{\rm d}\lambda &=&2d_1(\lambda
)g_1(g_2-g_1)+2d_2g_1g_4-2\,d_3g_1g_2,  \nonumber \\
{\rm d}g_2/{\rm d}\lambda &=&d_1(\lambda
)(g_2^2+g_3^2)+2d_2(g_1-g_2)g_4-d_3(g_1^2+g_2^2),  \nonumber \\
{\rm d}g_3/{\rm d}\lambda &=&-2d_0(\lambda )g_3g_4+2d_1(\lambda
)g_3(2g_2-g_1),  \nonumber \\
{\rm d}g_4/{\rm d}\lambda &=&-d_0(\lambda
)(g_3^2+g_4^2)+d_2(g_1^2+2g_1g_2-2g_2^2+g_4^2),  \label{TwoPatch}
\end{eqnarray}
where
\begin{eqnarray}
d_0(\lambda ) &=&2c_0\lambda ,\;d_2=2z_0;\;d_3=2c_{{\bf Q}};  \nonumber \\
d_1(\lambda ) &=&2\min (\lambda ,z_{{\bf Q}}).
\end{eqnarray}
Eqs. (\ref{TwoPatch}) have to be solved with the initial conditions $%
g_i(1)=g_i^0$. Note that we neglect here the renormalization of coupling
constants which arise from non-logarithmical contributions at $\lambda <1$,
this renormalization is small provided that the condition $g_i^0\ll 1$ is
satisfied.

Eqs. (\ref{TwoPatch}) coincide with the temperature cutoff RG approach for
one-particle irreducible functions on the patched FS \cite{SalmHon1} (see
Sect. IIIB) in the approximation that the vertex is constant (i.e. does not
depend on the patch) for the momenta in the vicinity of the vHS (for $|{\bf p%
}_i|<\Lambda $) and zero far from it (when $|{\bf p}_i|>\Lambda $ at least
for one $i=1...4$). Therefore, the real dependence of the vertices on the
momenta along the FS is replaced by a step function in the two-patch
approach. At the same time, the momentum dependence of the vertices in the
direction perpendicular to the FS is treated correctly through the scaling
variable $\lambda .$ Note however, that the momentum dependence of the
electronic spectrum within each patch is correctly taken into account in the
two-patch approach.

Strictly speaking, some of the two-loop diagrams (e.g. the self-energy
corrections) give contributions of the same order as the
single-logarithmical terms which we treat here and they should be taken into
account as well. However, the analysis of such contributions is rather
involved and the subject for future work \cite{Twoloop}. Note that this
difficulty is also ``hidden'' in the RG approaches on a patched FS where
additional logarithmical divergencies arise from the momentum integration
along the FS.

In order to explore the possible instabilities of the system, we consider
the behavior of the zero-frequency, time-ordered response functions at zero
temperature
\begin{equation}
\chi _m({\bf q})=\int\limits_{-\infty }^\infty {\rm d}\tau \langle T\,[%
\widehat{O}_m({\bf q},\tau )\widehat{O}_m(-{\bf q,}0)]\rangle
\end{equation}
where $\widehat{O}_m({\bf q},\tau )$ denote the Fourier components of the
operators (\ref{O}) in the Heisenberg representation. Besides the operators
in Eq. (\ref{O}) we also test, following Refs. \cite{Murakami,Binz}, the
susceptibilities which correspond to the operators
\begin{eqnarray}
\widehat{Q}(i) &=&\frac 12\sum_\sigma (c_{i\sigma }^{\dagger }c_{i\sigma
}-1),  \nonumber \\
\widehat{\tau }(i) &=&\frac 1{2z}\sum_{j\sigma }\lambda _{ij}c_{i\sigma
}^{\dagger }c_{j\sigma },  \nonumber \\
\widehat{A}(i) &=&\frac 1{2z}\sum_{j\sigma }\sigma \lambda _{ij}c_{i\sigma
}^{\dagger }c_{j\sigma },  \label{Op1} \\
\widehat{\pi }(i) &=&\frac{(-1)^i}{2z}\sum_{j\sigma }\lambda _{ij}c_{i\sigma
}^{\dagger }c_{j,-\sigma }^{\dagger },  \nonumber \\
\widehat{\eta }(i) &=&\frac{(-1)^i}{2z}\sum_{j(i),\sigma }\sigma c_{i\sigma
}^{\dagger }c_{j,-\sigma }^{\dagger },  \nonumber
\end{eqnarray}
where $j(i)$ are the nearest neighbor sites of site $i.$ The susceptibility $%
\chi _Q(0)$ is related to the isothermal compressibility $\kappa _T$ by
\begin{equation}
\kappa _T=\chi _Q(0)/n^2.  \label{kx}
\end{equation}
The bond-charge order parameter $\tau $ characterizes an instability towards
a spontaneous deformation of the FS\cite{Binz}, i.e. anisotropy in $x$ and $%
y $ directions, the so called Pomeranchuk instability \cite{Metz}.
Analogously, a bond-spin order parameter $A$ characterizes a spin-dependent
Pomeranchuk instability. The operators $\widehat{\pi }$ and $\widehat{\eta }$
correspond to triplet and singlet pairing with momentum ${\bf Q}=(\pi ,\pi )$%
.

Picking up the logarithmical divergences in $\lambda $ we obtain the RG
equations for the susceptibilities in the same approximations as discussed
above (cf. Refs. \cite{Schulz,Led,FurRice,OurVH})
\begin{eqnarray}
{\rm d}\chi _m(\lambda )/{\rm d}\lambda &=&d_{a_m}(\lambda ){\cal T}%
_m^2(\lambda ),  \label{hi} \\
{\rm d}\ln {\cal T}_m(\lambda )/{\rm d}\lambda &=&d_{a_m}(\lambda )\Gamma
_m(\lambda ),  \nonumber
\end{eqnarray}
where the coefficients $\Gamma _m$ are given by
\begin{equation}
\begin{array}{ll}
\Gamma _{{\rm CDW}}=g_2-g_3-2g_1; & \Gamma _{{\rm SDW}}=g_2+g_3; \\
\Gamma _{{\rm CF}}=g_2+g_3-2g_1; & \Gamma _{{\rm SF}}=g_2-g_3; \\
\Gamma _\pi =g_1-g_2; & \Gamma _\eta =-g_1-g_2; \\
\Gamma _Q=g_1-2g_2-g_4; & \Gamma _{{\rm F}}=g_1+g_4; \\
\Gamma _\tau =-g_1+2g_2-g_4; & \Gamma _A=g_4-g_1; \\
\Gamma _{{\rm dSC}}=g_3-g_4. &
\end{array}
\label{GG}
\end{equation}
In Eqs. (\ref{hi}) $a_m=0$ for the dSC phase; $a_m=1$ for SDW, CDW, SF, and
CF phases, $a_m=2$ for F, Q, $\tau $ and $A$ phases, and $a_m=3$ for $\pi $
and $\eta $ phases. Eqs. (\ref{hi}) have to be solved with the initial
conditions ${\cal T}_m(0)=1,$ $\chi _m(0)=0.$ At $\lambda =0$ the vertices (%
\ref{GG}) coincide with those considered in the mean-field approach in
section II (see Eq. (\ref{GMF})).

Numerical solutions of Eqs. (\ref{TwoPatch}) show, that at a critical value $%
\lambda _c$ of the scaling parameter $\lambda $ some of the vertices and
susceptibilities are divergent. We analyze the behavior of the coupling
constants $g_1$ to $g_4$ for $\lambda \rightarrow \lambda _c$ representing
it in the form ($\beta _1...\beta _4$) where $\beta _i=0,+$ or $-$ describes
the behavior of the coupling constant $g_i$: the ``plus'' (``minus'') sign
means relevant in the RG sense and tends to $+\infty $ ($-\infty $) and zero
means irrelevant. To identify the leading instabilities we calculate the
inverse susceptibilities for $\lambda <\lambda ^{*}$ ($\lambda ^{*}$ is the
value of the scaling parameter $\lambda $ when the largest absolute value of
the coupling constants $|g_i|$ exceeds unity) and then use a linear
extrapolation of the inverse susceptibility (see Fig. 7 below). From the RG
point of view, at $\lambda \sim \lambda ^{*}$ the ``one--dimensional-like''
behavior of the coupling constants considered above changes to
two-dimensional and also the parts of the FS far from the van Hove points
become non-negligible. However, since the values of the susceptibilities are
large already at $\lambda ^{*},$ the corresponding critical region is narrow.

For a given $\lambda _c$ the size $\Lambda $ of the patches is restricted by
$\Lambda \ll \pi $ and $\,\ln (4/\Lambda )\ll \lambda _c.$ The latter
criterion follows from the condition that the contribution of the electrons
with $|k_{\pm }|<\Lambda $ to particle-hole and particle-particle bubbles is
dominant (see, e.g. Ref. \cite{Binz}). We choose $\Lambda =1$ and require $%
\lambda _c\gg \ln 4\simeq 1.$ Since $\lambda _c$ decreases with increasing
interaction, this criterium restricts the values of the interactions where
the two-patch RG approach is valid.

\subsection{Many-patch analysis}

In the many-patch analysis we follow the temperature-cutoff renormalization
group for one-particle irreducible (1PI) Green functions proposed recently
by Honerkamp and Salmhofer in Ref. \cite{SalmHon1}. This version of the RG
uses the temperature as a natural cutoff parameter, allowing to account for
both the excitations with momenta close to the Fermi surface and far from
it, which is necessary for the description of the particle-hole
instabilities with zero momentum transfer, e.g. ferromagnetism, phase
separation, and the Pomeranchuk instability. Neglecting the frequency
dependence of the vertices, which is justified in the weak-coupling regime,
the RG differential equation for the temperature- and momentum-dependent
electron-electron interaction vertex (see the diagrammatic representation in
Fig. 3), has the form
\begin{eqnarray}
&&\ \ \ \ \frac d{dT}V_T({\bf k}_1,{\bf k}_2,{\bf k}_3)
\begin{array}{c}
=
\end{array}
-\sum_{{\bf p}}V_T({\bf k}_1,{\bf k}_2,{\bf p})L_{pp}({\bf p},-{\bf p}+{\bf k%
}_1+{\bf k}_2)V_T({\bf p},-{\bf p}+{\bf k}_1+{\bf k}_2,{\bf k}_3)  \nonumber
\\
&&\ \ \ \ \ \ \ \;\;+\sum_{{\bf k}}\big[ -2V_T({\bf k}_1,{\bf p},{\bf k}%
_3)V_T({\bf p}+{\bf k}_1-{\bf k}_3,{\bf k}_2,{\bf p})+V_T({\bf k}_1,{\bf p},%
{\bf k}_3)V_T({\bf k}_2,{\bf p}+{\bf k}_1-{\bf k}_3,{\bf p})  \nonumber \\
&&\ \ \ \ \ \ \ \;\;+V_T({\bf k}_1,{\bf p},{\bf p}+{\bf k}_1-{\bf k}_3)V_T(%
{\bf p}+{\bf k}_1-{\bf k}_3,{\bf k}_2,{\bf p})\big] L_{ph}({\bf p},{\bf p}+%
{\bf k}_1-{\bf k}_3)  \nonumber \\
&&\ \ \ \ \ \ \ \;\;+\sum_{{\bf k}}V_T({\bf k}_1,{\bf p}+{\bf k}_2-{\bf k}_3,%
{\bf p})L_{ph}({\bf p},{\bf p}+{\bf k}_2-{\bf k}_3)V_T({\bf p},{\bf k}_2,%
{\bf k}_3)  \label{dV}
\end{eqnarray}
where
\begin{eqnarray}
L_{ph}({\bf k},{\bf k}^{\prime }) &=&\frac{f_T^{\prime }(\varepsilon _{{\bf k%
}})-f_T^{\prime }(\varepsilon _{{\bf k}^{\prime }})}{\varepsilon _{{\bf k}%
}-\varepsilon _{{\bf k}^{\prime }}}  \nonumber \\
L_{pp}({\bf k},{\bf k}^{\prime }) &=&\frac{f_T^{\prime }(\varepsilon _{{\bf k%
}})+f_T^{\prime }(\varepsilon _{{\bf k}^{\prime }})}{\varepsilon _{{\bf k}%
}+\varepsilon _{{\bf k}^{\prime }}}
\end{eqnarray}
and $f_T^{\prime }(\varepsilon )=df(\varepsilon )/dT.$ Eq. (\ref{dV}) has to
be solved with the initial condition $V_{T_0}({\bf k}_1,{\bf k}_2,{\bf k}%
_3)=g({\bf k}_1,{\bf k}_2,{\bf k}_3,{\bf k}_1+{\bf k}_2-{\bf k}_3)$ where
the initial temperature $T_0$ is of the order of the bandwidth$.$ The
evolution of the vertices with decreasing temperature determines the
temperature dependence of the susceptibilities according to\cite
{SalmHon1,Hon2002}
\begin{eqnarray}
\frac d{dT}\chi _{mT} &=&\sum_{{\bf k}^{\prime }}{\cal T}_{mT}({\bf k}%
^{\prime }){\cal T}_{mT}(\mp {\bf k}^{\prime }+{\bf Q}_m)L_{pp,ph}({\bf k}%
^{\prime },\mp {\bf k}^{\prime }+{\bf q}_m),  \label{dH} \\
\frac d{dT}{\cal T}_{mT}({\bf k}) &=&\mp \sum_{{\bf k}^{\prime }}{\cal T}%
_{mT}({\bf k}^{\prime })\Gamma _{mT}({\bf k},{\bf k}^{\prime })L_{pp,ph}(%
{\bf k}^{\prime },\mp {\bf k}^{\prime }+{\bf q}_m)  \nonumber
\end{eqnarray}
where
\begin{equation}
\Gamma _{mT}({\bf k,k}^{\prime })=\left\{
\begin{array}{cl}
V_T({\bf k},{\bf k}^{\prime },{\bf k}^{\prime }+{\bf q}_m)-2V_T({\bf k},{\bf %
k}^{\prime },{\bf k}+{\bf q}_m) & \text{for }m=\text{CDW, CF, Q and }\tau ,
\\
V_T({\bf k},{\bf k}^{\prime },{\bf k}^{\prime }+{\bf q}_m) & \text{for }m=%
\text{SDW,SF,F, and A,} \\
V_T({\bf k},-{\bf k+q}_m,{\bf k}^{\prime }) & \text{for }m=\pi ,\eta ,\text{
and dSC.}
\end{array}
\right.
\end{equation}
${\bf q}_m={\bf Q}$ for CDW, SDW, CF, SF, $\pi $ and $\eta ,$ and ${\bf q}_m=%
{\bf 0}$ otherwise. The upper signs in Eq. (\ref{dH}) refer to the
particle-particle response ($\pi ,\eta $ and dSC), the lower signs to the
particle-hole response. The initial conditions for Eqs. (\ref{dH}) are
\begin{equation}
{\cal T}_{m,T_0}({\bf k})=\left\{
\begin{array}{cl}
\cos k_x-\cos k_y & \text{for CF, SF, PI, A, dSC, and }\pi , \\
1\text{ } & \text{otherwise,}
\end{array}
\right.   \label{T}
\end{equation}
and $\chi _{m,T_0}=0.$ To solve numerically Eqs. (\ref{dV}) and (\ref{dH}),
we use the discretization of momentum space in 32 patches and the same
patching scheme as proposed in Ref. \cite{SalmHon1} (we have checked in
selected cases that increasing the number of patches to 48 does not change
the results). With account of the symmetries of the square lattice, this
reduces the above integro-differential equations to a set of 1920
differential equations which were solved numerically. We use the value of
the starting temperature $T_0=12t,$ which is slightly larger than the
bandwidth, and stop the flow of the coupling constants where the largest
coupling constant $V_{\max}=18t$. As for the two-patch analysis, we
extrapolate the resulting inverse susceptibilities and calculate the critical
values of the scaling parameter $\lambda _c^m$ where the extrapolated
inverse susceptibilities vanish. Due to this procedure, the results of RG
analysis do not depend strongly on $V_{\max }.$ Note that the initial
$k$-dependence of the response functions (\ref{T}) is slightly changed
during the renormalization-group flow: responses with $d$-wave symmetry  (CF,
SF, PI, A, dSC, and $\pi $) acquire $g$-wave and higher-order harmonics,
while responses with $s$-wave symmetry acquire additional extended $s$-wave
($\cos k_x+\cos k_y$). However, these additional corrections are small.

\subsection{Results of the RG analysis}

For the calculations we choose the interaction strength $U=2t$ since for
stronger interactions and moderate $|J|/U$ and $|V|/U$ the RG approach
becomes unreliable. The phase diagrams calculated with this value of $U$,
different values of $t^{\prime }/t=0,$ $0.1,$ $0.3\ $and the corresponding
van Hove fillings are presented in Figs. 4-6. The phase boundaries,
determined from the two-patch approach are shown by thin solid and dashed
lines. Solid lines separate the phases with different behavior of the
coupling constants, while dashed lines separate phases with the same
behavior of the coupling constants and are determined from the condition of
the equality of the corresponding critical values $\lambda =\lambda _c^m$
where the linearly extrapolated susceptibilities vanish.

We consider first the results of the two-patch approach. Since this approach
is applicable for not too small $\lambda _c$ (see Section IIIA), we consider
only the region of the phase diagrams with $\lambda _c>2,$ which is bounded
by bold lines. In the most part of the phase diagrams the typical values of $%
\lambda _c$ range from $3$ to $5.$ The dependencies of the inverse
susceptibilities on the scaling parameter for selected parameter values are
shown in Fig. 7. For not too large values of $|J|$ ($J<0$) and for $%
t^{\prime }=0$ or $t^{\prime }=0.1t$ (Figs. 4,5) the two-patch approach
predicts that the spin-flux phase of the type $(0+--)$ is the leading
instability. With increasing $V$ this phase is replaced by a phase which has
comparable CDW and CF susceptibilities (CDW phase) and the behavior of the
coupling constants $(--0+)$. With increasing $J$ up to $J\gtrsim 0.25U$ the
CDW susceptibility dominates but also $\chi _\tau $ and $\chi _{SF}$ are
large (we term this phase CDW$^{\prime }$). The corresponding flow of the
coupling constants is $(0+--).$ At large negative $V$ a phase of type $(0-0-)
$ occurs where the largest diverging susceptibility is $\chi _Q$ which,
according to Eq. (\ref{kx}), implies a divergent isothermal compressibility $%
\kappa _T.$ A divergence of $\kappa _T$ can be attributed generally to
different phenomena: a metal-insulator transition \cite{MIT} or phase
separation. In the considered parameter region the most natural explanation
is phase separation induced by a large negative $V.$ For large enough $|J|$ $%
(J<0)$ we obtain the ferromagnetic state (F) with the coupling constant flow
$(++0+)$; for large positive $V$ $\chi _A$ is the most diverging
susceptibility. However, for small $t^{\prime }\lesssim 0.3t$ these
instabilities are outside of the weak-coupling region of the phase diagram.
The dashed area corresponds to the frustrated regime where the critical
value of the scaling parameter $\lambda _c>20$ and the spin-density-wave
instability is strongly suppressed by negative $J.$ Such behavior arises
from the competition between antiferromagnetic and superconducting
fluctuations on one side and ferromagnetic fluctuations on the other side.
As a result different types of instabilities almost ``cancel'' each other.

Four other phases (SDW, dSC, CF and PI) have the same flow of the coupling
constants ($0++-$). The magnitudes of the susceptibilities $\chi _{{\rm dSC}%
},$ $\chi _{{\rm SDW}},$ $\chi _{{\rm CF}}$ in the region $J>0$ are close to
each other (see Fig. 7b), which naturally implies a close competition
between these states. With increasing $J$ we find a PI phase where the
susceptibility $\chi _\tau $ is largest. However, this instability also
always appears outside of the weak-coupling region of the phase diagram
where the equations (\ref{TwoPatch}) are valid.

As in the mean-field solution, away from half filling one expects phase
separation or the formation of inhomogeneous structures of all ordered
phases with wavevector ${\bf Q}$ (CDW, SDW, CF and SF). Note however, that
this filling-induced type of phase separation should be contrasted with the
interaction-induced phase separation in the PS region of the phase diagram,
which is present even at half-filling ($t^{\prime }=0$) and may not be
magnetically ordered or superconducting.

The results of the many-patch approach are shown by different symbols,
explained in the figure captions. Solid symbols correspond to different
types of ordering tendencies. Open symbols correspond to frustrated
behavior, where the many-patch RG could not reach $V_{\max }=18t$ for $%
\lambda _T=\ln (t/T)/2<4.$ Boundaries between some phases (e.g. CDW and SF,
dSC and PS for $J>0$) almost coincide in the two- and many-patch approaches.
At the same time, the many-patch approach predicts a much broader region of
stability for the SDW phase and an almost vanishing region of stability for
the CF phase. Therefore, for $t^{\prime }=0$ and $t^{\prime }=0.1$ taking
into account more patches on the FS (and therefore the contribution of
nested parts) moves the predicted phase boundaries closer to those of the
mean-field results. At the same time, the charge flux, as well as the $d$%
-wave superconducting response in the many-patch approach are substantial
within the antiferromagnetic phase, which implies a close competition
between the above ordering tendencies. For a rough estimate, where the
charge response is closest to the antiferromagnetic one, we bound the region
with $\chi _{\text{CF}}>$ $(2/3)\chi _{\text{SDW}}$ by thin crosses.

At $J<0$ many-patch approach does confirm the possibility of a spin flux
phase in almost the same parameter region as determined from the two-patch
approach. With decreasing $V$ this spin flux phase is however replaced by a
(partially frustrated) ferromagnetic phase, which fills most of the region
where the frustrated SDW order was expected in the two-patch approach. The
presence of the ferromagnetic phase at $J<0$ resembles closeby the
mean-field prediction in this parameter range. The $d$-wave superconducting
region for $J<0$ is replaced by phase separation in the many-patch approach.

Note that with decreasing of $U$ the region of $d$-wave superconductivity at
$J>0$ grows, while increasing of $U$ favors the SDW instability at $J>0$.
Similarly, at $J<0$ decreasing $U$ shifts the balance between SF phase
and ferromagnetism towards the spin-flux phase. Other phases (e.g. CDW and
PS) are essentially not influenced by varying $U.$

When increasing $t^{\prime }$ up to $t^{\prime }=0.3t$ the phase diagram
changes (see Fig. 6). In the two-patch approach, at $J>0.05U$ $d$-wave
superconductivity becomes the leading instability. For large positive $V$ we
obtain either CDW and A instabilities (CDW phase on Fig. 6) or CDW and PI
instabilities (CDW$^{\prime }$ phase) depending on $J$. For large negative $%
V $ we again find interaction-induced phase separation. Most of the region
with $J<0.05U$ and moderate $|V|$ is frustrated ($\lambda _c>20$), which is
again the result of a strong competition of different ordering tendencies.

The many-patch approach also gives the strongest tendency to $d$-wave
superconductivity at $J>0.05U,$ but the tendency towards ferromagnetic order
in a large part of the frustrated region $J<0.05U$. Similar to $t^{\prime
}=0 $ and $t^{\prime }=0.1t,$ at $t^{\prime }=0.3t,$ the $d$-wave
superconducting region for $J<0$ predicted by the two-patch approach is
replaced by phase separation in the many-patch approach. The boundary of the
CDW phase also almost coincides for the many-patch and two-patch approach.
Therefore, at $t^{\prime }=0.3t$ the predictions of the two-patch approach
are closer to the results of the many-patch approach, than for smaller $%
t^{\prime }$. This is connected with the absence of nested parts of the
Fermi surface, as discussed above.

In summary, the RG analysis gives a much richer phase diagram than
anticipated from the mean-field results of the previous section. At the same
time, for small $t^{\prime }$ the results of the many-patch RG approach,
which takes into account the contributions of the whole FS and therefore
treats the renormalization of the interactions in a more accurate way, are
closer to the mean-field predictions than the results of the two-patch
approach. At intermediate $t^{\prime }$ the two-patch approach becomes more
reliable and the results of both approaches are close. In the next section
we supplement the above RG studies by an $SO(8)$ symmetry analysis.

\section{Symmetries and asymptotic behavior}

Additional insight to the phase diagram is obtained by considering the
symmetries of the Hamiltonian (\ref{H}) with respect to the $SO(8)$ symmetry
group of transformations of the $4$-fermion states $|{\bf k}\uparrow \rangle
,|{\bf k}\downarrow \rangle ,|{\bf k}+{\bf Q}\uparrow \rangle $ and $|{\bf k}%
+{\bf Q}\downarrow \rangle $ (see e.g. Ref.\cite{Markiewicz}). This group is
generated by the $so(8)$ algebra of 28 operators which are given in part by
\begin{equation}
\widehat{O}_m=\frac 1N\sum_i\widehat{O}_m(i)
\end{equation}
with $\widehat{O}_m(i)$ as defined above in Eqs. (\ref{O}) and (\ref{Op1}).
For a full list of operators of the $so(8)$ algebra we refer to Ref. \cite
{Markiewicz}. From the operators $\widehat{O}_m$ of the $so(8)$ algebra we
construct the rotation operators
\begin{equation}
R_m(\alpha )=\exp (\text{i}\alpha \widehat{O}_m^{\dagger }+\text{i}\alpha
\widehat{O}_m)
\end{equation}
where $\alpha $ is a real number. From the symmetry point of view, the
operators $\widehat{O}_m=\widehat{\pi },\widehat{\eta },\widehat{A},$ and $%
\widehat{\tau }$ are most useful. As discussed in Refs. \cite
{Markiewicz,Murakami}, the operator $R_\pi $ performs a rotation between SDW
and dSC states (which is the basis of the $SO(5)$ theory \cite{SO5}), the
operator $R_\eta $ between dSC and CF states, the operator $R_A$ between SDW
and CF phases as well as CDW and SF phases while the operator $R_\tau $
rotates between SDW and SF phases and between CDW and CF phases,
respectively.

The symmetry of the Hamiltonian (\ref{H}) under these operations is as
follows. The non-interacting part of the Hamiltonian is invariant ($%
H_0=R_mH_0R_m^{-1}$) under the rotations $R_A$ and $R_\tau $ and in the
nesting case $\varepsilon _{{\bf k}}=-\varepsilon _{{\bf k+Q}}$ under $R_\pi
$ and $R_\eta .$ The interaction is invariant only under the rotations $%
R_\eta $ for $V=J=0,$ i.e. the Hubbard model, which was originally
discovered by Yang and Zhang \cite{YangZhang} but the interaction is not
invariant with respect to the other rotation operations.

The symmetry of the restricted two-patch Hamiltonian (\ref{HVH}) is
considerably higher\cite{Binz}. For this restricted Hamiltonian the
interaction part, and therefore the entire Hamiltonian at half-filling and
zero $t^{\prime },$ is invariant with respect to the operations $R_\pi
,R_\eta ,R_A$ or $R_\tau $ on special lines in the $V$-$J$ plane (at fixed $%
U $). The corresponding lines are supplied by captions in Fig. 4,
corresponding to the type of the symmetry. The symmetry lines exactly
coincide with the boundaries between different phases in the two-patch
approach, since the corresponding susceptibilities $\chi _m(\lambda )$ in
this approach are identical on these lines for arbitrary $\lambda $. At the
same time, as we have seen in Section III, most of these symmetries (those
which are connected with the SDW order) are broken by the contribution of
the nested parts of FS. Away from half-filling or at finite $t^{\prime }$
the non-interacting part of the Hamiltonian becomes invariant only with
respect to the rotations $R_A$ and $R_\tau $.

To get more insight into the posibility of different ordering tendencies, we
rewrite the Hamiltonian in terms of the operators of the $so(8)$ algebra. We
restrict ourselves to the contribution of the vH points, since the
corresponding analysis for the whole FS becomes very complicated and
requires numerical diagonalization of the corresponding Hamiltonian. Taking
into account that the non-interacting part of the Hamiltonian (\ref{HVH})
for vH points, i.e.
\begin{equation}
H_0=\varepsilon _{{\bf k}_A}a_{{\bf 0},\sigma }^{\dagger }a_{{\bf 0},\sigma
}+\varepsilon _{{\bf k}_B}b_{{\bf 0},\sigma }^{\dagger }b_{{\bf 0},\sigma }
\end{equation}
\ vanishes at the vH filling (where $\varepsilon _{{\bf k}_A}=\varepsilon _{%
{\bf k}_B}=0$), the Hamiltonian is rewritten as
\begin{eqnarray}
H_{eff} &=&g_1(\widehat{O}_{{\rm CDW}}^2-\widehat{O}_{{\rm CF}}^2)\,+\,g_2(%
\widetilde{Q}^2-\tau ^2)+g_3(\widehat{O}_{{\rm CDW}}^2+\widehat{O}_{{\rm CF}%
}^2)  \label{Heff} \\
&&\ \ +g_4(\widetilde{Q}^2+\tau ^2)-2\widetilde{Q}(g_2+g_4)  \nonumber
\end{eqnarray}
where $\widetilde{Q}=\widehat{Q}+1/2$ and $g_1$ ... $g_4$ are the vertices
determined in the two-patch approach (Sect. IIIA). These vertices can be
also considered as those obtained from the many-patch analysis for the
corresponding momenta, provided that the contribution of electrons with
momenta far from the vH singularities is small, i.e. $t^{\prime }/t$ is not
close to zero$.$ Under these conditions, the Hamiltonian (\ref{Heff}) can be
considered as an effective Hamiltonian of the RG procedure in the previous
section, which acts on the space of $16$ states $|s_A,s_B\rangle $ where $%
s_{A,B}=0,\uparrow ,\downarrow $ or $\uparrow \downarrow $ denotes the
electron states with vH momenta ${\bf k}_A$ or ${\bf k}_B$.

First we consider the eigenstates $|m\rangle $ of the operators $\widehat{O}%
_m,$ which are easily expressed in terms of the states $|s_A,s_B\rangle $
and we obtain the expectation values of the Hamiltonian in the corresponding
states as
\begin{equation}
E_m(\lambda )=\frac{\langle m|H_{eff}|m\rangle }{\langle m|m\rangle }%
=C(\lambda )-\Gamma _m(\lambda )/2
\end{equation}
where $C(\lambda )=-(g_1+2g_2-2g_3+g_4)/2$ is independent of $m$ and $\Gamma
_m(\lambda )$ is defined in Eq. (\ref{GG}). Therefore the states which were
identified in the two-patch approach as having the largest susceptibility
(largest $\Gamma _m(\lambda )$ for $\lambda \rightarrow \lambda _c$) have
also the lowest effective energy among the eigenstates $|m\rangle $.

For $m\neq {\rm F}$ or ${\rm Q}$ the eigenstates $|m\rangle $ of the
operators $\widehat{O}_m$ are not the eigenstates of the effective
Hamiltonian (\ref{Heff}), since the operators $\widehat{O}_m$ generally do
not commute with this Hamiltonian. To find the eigenstates of the
Hamiltonian, we represent it as a matrix with respect to the states $%
|s_A,s_B\rangle $ and diagonalize it. In fact, the subspaces with even (odd)
number of particles
\begin{equation}
N_{VH}=\langle a_{{\bf 0},\sigma }^{\dagger }a_{{\bf 0},\sigma }+b_{{\bf 0}%
,\sigma }^{\dagger }b_{{\bf 0},\sigma }\rangle
\end{equation}
are not mixed by the operators of the $so(8)$ algebra and we can consider
them separately. In the following, we focus on the even subspace where the
interaction part of the Hamiltonian is non-trivial. The resulting energy
levels and the eigenfunctions are presented in Table 1. The states $%
|V_i\rangle $ are the triplet of eigenvectors of the operator $\widehat{O}_{%
{\rm F}}$ with eigenvalues $1,0$ and $-1,$ respectively. From the states $%
|U_i\rangle $ one can form a doublet of vectors $|U_1\rangle \pm |U_2\rangle
$ which are the eigenvectors of the operator $\widehat{Q}$ with eigenvalues $%
\mp 1$. The other states\ are not eigenvectors of any operator in Eqs. (\ref
{O}) and (\ref{Op1}). Considering the flow of the energy levels of the
respective states which are given as a function of the coupling constants $%
g_i$ in the second column of Table 1, we obtain the correspondence between
the lowest energy eigenstates of the effective Hamiltonian in the two-patch
RG procedure and the phases identified in Sect. III. We find that in the F
and SF phases the triplet $|V_i\rangle $ has the lowest energy. It is
natural to associate the states $|V_{1,3}\rangle ,$ which have eigenvalues $%
\pm 1$ of the operator $\widehat{O}_{{\rm F}},$ with the ferromagnetic phase
and the state $|V_2\rangle $ with the SF phase. Similarly, in the PS phase
the states $|U_i\rangle $, in the CDW and A phases the state $|C_1\rangle ,$
and in the CDW$^{\prime }$ phase the state $|C_2\rangle $ have the lowest
energies. Finally, the $|W\rangle $ state is common for SDW, CF, dSC, and PI
instabilities, i.e. phases which have the coupling constants flow of ($0++-$%
) type. The resulting correspondence is presented in the right column of
Table 1.

To clarify possible types of orders which correspond to the eigenstates $%
|k\rangle $ of the effective Hamiltonian (\ref{Heff}), we expand these
eigenstates in terms of the eigenvectors $|m\rangle $ of the operators $%
\widehat{O}_m$ by calculating the scalar products $C_{mk}=\langle k|m\rangle
$ listed in Table 2. Generally the states $|k\rangle $ are a mixture of the
states which correspond to different order parameters. So, the state $%
|C_1\rangle $ in Table 1, which was identified above with CDW and A phases,
mixes CDW, CF, $A,$ and $\eta $ types of order. The state $|C_2\rangle $
(identified with the CDW$^{\prime }$ phase) mixes CDW, PI, and SF types of
order. The state $|U_2\rangle $ which is connected with the PS phase is also
a mixture of dSC and $\eta $ pairing. Finally the $|W\rangle $ state mixes
SDW, dSC, CF, and PI types of order in complete agreement with the results
for the susceptibilities in the two-patch approach.

Although we do not analyze in this section the contribution of the whole FS,
one can expect that general features learned from the simple two-patch
analysis hold in this case too. Namely, we expect that the eigenstates of
the whole Hamiltonian do not coincide with the eigenstates of operators,
corresponding to different order parameters. Although the eigenvectors of
the effective Hamiltonian corresponding to the contribution of the whole FS
have a much more complicated form, we expect that the combinations of the
order parameters listed in Table 2 which belong to the {\it same }eigenstate
remain unchanged. This can be also seen from the analysis of simultaneously
diverging susceptibilities in the many-patch approach. At the same time, one
can not safely argue whether or not the discussed states support the
coexistence of long-range orders, or whether part of the abovementioned
orders are only short-range. Resolving these possibilities requires a
strong-coupling analysis of the problem which is beyond the validity of
one-loop renormalization-group approach.

\section{Summary and conclusions}

We considered the phase diagrams of the extended $U$-$V$-$J$ Hubbard model
within mean-field approximation and from RG approaches. The extended Hubbard
model has a very complex phase diagram with various types of orders. The
mean-field approximation provides only a rough phase diagram of the system.
At positive $J$ it leads mainly to the coexistence of antiferromagnetism
with a small amplitude of $d$-wave superconductivity. At negative $J$
mean-field theory predicts ferromagnetic order. For large enough positive $V$
these phases are replaced by CDW order, while for large negative $V$ the
ground state is a $d$-wave superconductor. Charge- and spin- flux phases, as
well as their coexistence with $d$-wave superconductivity are never
energetically stable within the mean-field approximation.

The RG analyses at van Hove band fillings leads to substantially reacher
phase diagrams (Figs. 4-6). Instabilities towards SDW, CDW, dSC, CF, SF, F
order are possible in different parameter regimes. Importantly, for $J\geq 0$
and not too large $|V|$ SDW, CF, dSC orders have comparable
susceptibilities, which signals their close competition in this parameter
region. At the same time, the tendency towards the formation of charge-flux
is greatly suppressed in the many-patch approach in comparison with the
two-patch results, and it becomes the leading instability only in the
restricted parameter range. With increasing $|t^{\prime }|$ the $d$-wave
superconducting state becomes more preferable. Tendencies towards both
spin-independent and spin-dependent Pomeranchuk instabilities are outside of
the weak-coupling region of the phase diagrams where the RG analysis is
applicable.

The symmetry analysis shows that most of the phases obtained within the RG
approach are in fact a mixture of long- or short-range orders of different
types. Resolving between short- and long-range orders for these phases
requires a strong-coupling analysis of the problem which can not be
performed within the RG approach.

Comparing the obtained results with the phase diagram of the 1D version of
the $U$-$V$-$J$ Hubbard model \cite{UVJ1D1}, we observe that the phase
diagram of 1D model is similar to the results of the two-patch approach at
half-filling. Charge- and spin-flux phases replace the staggered
bond-order-wave phase of the 1D model. The CDW, CDW$^{\prime }$ phases
appear in two dimensions in the same parameter region as the CDW phase in
the 1D case. For $J>0$ the SDW phase of the 1D system is partially
substituted by $d$-wave superconductivity. At the same time, the results of
two-patch approach are strongly changed at half filling by the contribution
of the whole Fermi surface and become closer to the mean-field phase
diagram, rather than to 1D results.

The physically most relevant regime of the 2D $U$-$V$-$J$ Hubbard model is
contained in the parameter range $0<V<U/4,$ $0<J<U/2$. The corresponding
areas are shaded in Figs. 4-6. With decreasing van Hove filling we observe
the following qualitative changes in the shaded area of the phase diagram:
at half-filling SDW order dominates. For small doping $\delta =1-n_{VH}=0.08$%
, we encounter a very complicated situation, in which SDW, charge-flux and $%
d $-wave superconductivity ordering tendencies are all simultaneously
strong, although SDW order still dominates. In this regime the ground state
structure is very sensitive to small parameter changes leading to a variety
of possible phase transitions and the possibility for coexisting phases.
Finally, at larger doping $\delta =0.28$ the picture becomes simpler again
and only $d$-wave superconductivity becomes the leading instability.

Under the abovementioned circumstances of strong competition of different
order parameters, the self-energy effects which are not accounted for in the
one-loop RG approach can become crucial. Therefore, for a final conclusion
about the possibility of a charge-flux phase in a broad parameter range of
the one-band $U$-$V$-$J$ model, the analysis of the two-loop contributions
to the RG schemes remains to be performed. Another topic for future work will
be to investigate to what extent the results for the one-band model $U-V-J$
model in a specific parameter range allow implications for the competing
orders in cuprate materials.

\section{Acknowledgements}

We are grateful to G. I. Japaridze, T. M. Rice, W. Metzner, and B. Binz for
valuable discussions and M. Vojta for important comments. This work was
supported by the Deutsche Forschungsgemeinschaft through SFB 484.

\newpage
\
\[
{\sc Tables}
\]

\[
\begin{tabular}{||lcll||}
\hline\hline
\multicolumn{1}{||l||}{$k$} & \multicolumn{1}{c|}{$E_k(\{g_i\})$} &
\multicolumn{1}{l|}{State} & Corresp. phases \\ \hline\hline
\multicolumn{1}{||l||}{1.} & \multicolumn{1}{c|}{$-g_2-g_4$} &
\multicolumn{1}{l|}{$\,|V_1\rangle =|\uparrow ,\uparrow \rangle
,\;|V_2\rangle =|\uparrow ,\downarrow \rangle +|\downarrow ,\uparrow \rangle
,\;|V_3\rangle =|\downarrow ,\downarrow \rangle $} & F, SF \\ \hline
\multicolumn{1}{||l||}{2.} & \multicolumn{1}{c|}{$0$} & \multicolumn{1}{l|}{$%
\,|U_1\rangle =|0,0\rangle +|\uparrow \downarrow ,\uparrow \downarrow
\rangle ;\;|U_2\rangle =|0,0\rangle -|\uparrow \downarrow ,\uparrow
\downarrow \rangle $} & PS \\ \hline
\multicolumn{1}{||l||}{3.} & \multicolumn{1}{c|}{$\,2g_1-g_2-g_4$} &
\multicolumn{1}{l|}{$\,|C_1\rangle =|\uparrow ,\downarrow \rangle
-|\downarrow ,\uparrow \rangle \;$} & CDW,\ A \\ \hline
\multicolumn{1}{||l||}{4.} & \multicolumn{1}{c|}{$\,g_1-2g_2+g_3$} &
\multicolumn{1}{l|}{$\,|C_2\rangle =|\uparrow \downarrow ,0\rangle
+|0,\uparrow \downarrow \rangle $} & CDW$^{\prime }$ \\ \hline
\multicolumn{1}{||l||}{5.} & \multicolumn{1}{c|}{$\,g_1-2g_2-g_3$} &
\multicolumn{1}{l|}{$\,|W\rangle =-|\uparrow \downarrow ,0\rangle
+|0,\uparrow \downarrow \rangle $} & SDW,\ CF,\ dSC,\ PI \\ \hline\hline
\end{tabular}
\]
Table 1. The eigenvalues $E_k(\{g_i\})$ and eigenstates $|k\rangle $ of the
effective Hamiltonian, Eq. (\ref{Heff}). The right column represents the
correspondence with the phases in the phase diagrams (see text).

\[
\begin{tabular}{||ccccc|cc|cc||}
\hline\hline
\multicolumn{1}{||c||}{State} & \multicolumn{1}{c|}{$\,|${\rm SDW}$\rangle $}
& \multicolumn{1}{c|}{$\,|{\rm dSC}\rangle $} & \multicolumn{1}{c|}{$|{\rm CF%
}\rangle $} & \multicolumn{1}{c|}{$|{\rm SF}\rangle $} &
\multicolumn{1}{|c|}{$\,|{\rm CDW}\rangle $} & \multicolumn{1}{c|}{$|\tau
\rangle $} & \multicolumn{1}{|c|}{$|A\rangle $} & $|\eta \rangle $ \\
\hline\hline
\multicolumn{1}{||c||}{$\langle V_2|$} & \multicolumn{1}{c|}{$1/\sqrt{2}$} &
\multicolumn{1}{c|}{0} & \multicolumn{1}{c|}{0} & \multicolumn{1}{c|}{$i/%
\sqrt{2}$} & \multicolumn{1}{|c|}{0} & \multicolumn{1}{c|}{0} &
\multicolumn{1}{|c|}{$1/\sqrt{2}$} & 0 \\ \hline
\multicolumn{1}{||c||}{$\langle C_1|$} & \multicolumn{1}{c|}{0} &
\multicolumn{1}{c|}{0} & \multicolumn{1}{c|}{$-i/\sqrt{2}$} &
\multicolumn{1}{c|}{0} & \multicolumn{1}{|c|}{$-1/\sqrt{2}$} &
\multicolumn{1}{c|}{0} & \multicolumn{1}{|c|}{$-1/\sqrt{2}$} & $-1/\sqrt{2}$
\\ \hline
\multicolumn{1}{||c||}{$\langle C_2|$} & \multicolumn{1}{c|}{0} &
\multicolumn{1}{c|}{0} & \multicolumn{1}{c|}{0} & \multicolumn{1}{c|}{$1/%
\sqrt{2}$} & \multicolumn{1}{|c|}{$1/\sqrt{2}$} & \multicolumn{1}{c|}{$1/%
\sqrt{2}$} & \multicolumn{1}{|c|}{0} & 0 \\ \hline
\multicolumn{1}{||c||}{$\langle U_2|$} & \multicolumn{1}{c|}{0} &
\multicolumn{1}{c|}{$-1/\sqrt{2}$} & \multicolumn{1}{c|}{0} &
\multicolumn{1}{c|}{0} & \multicolumn{1}{|c|}{0} & \multicolumn{1}{c|}{0} &
\multicolumn{1}{|c|}{0} & $-1/\sqrt{2}$ \\ \hline
\multicolumn{1}{||c||}{$\langle W\,|$} & \multicolumn{1}{c|}{$1/\sqrt{2}$} &
\multicolumn{1}{c|}{$-1/\sqrt{2}$} & \multicolumn{1}{c|}{$1/\sqrt{2}$} &
\multicolumn{1}{c|}{0} & \multicolumn{1}{|c|}{0} & \multicolumn{1}{c|}{$-1/%
\sqrt{2}$} & \multicolumn{1}{|c|}{0} & 0 \\ \hline\hline
\end{tabular}
\]
Table 2. The scalar products $C_{mk}=\langle k|m\rangle $ between the
eigenstates of the effective Hamiltonian, Eq. (\ref{Heff}) (see also Table
1) and the eigenstates of the operators in Eqs. (\ref{O}) and (\ref{Op1}).

\newpage\
\[
{\sc Figure\ captions}
\]

Fig. 1. The mean-field phase diagram for $U=4t,$ $t^{\prime }/t=0.1$ and van
Hove band filling $n=0.92.$ SDW and CDW denote the spin- and charge-density
wave phases, dSC the $d$-wave superconducting phase, SDW+$d$SC marks
phase coexistence, and F is the ferromagnetic phase. Solid and dashed lines
correspond to first and second order transitions, respectively.

Fig. 2. The Fermi surface at van Hove band fillings: $t^{\prime }=0$ and $%
n=1 $ (solid line), $t^{\prime }/t=0.1$ and $n=0.92$ (long-dashed line),$\ $%
and $t^{\prime }/t=0.3$ and $n=0.72$ (short-dashed line), $A$ and $B$ are
van Hove points.

Fig. 3. Diagrammatic representation for the many-patch RG equations, Eq. (\ref
{dV}). Lines drawn through the vertices show the direction of spin
conservation. Diagrams are drawn in the same order as the respective terms in
Eq. (\ref{dV}). The cutting dash at the propagator lines means the derivative
with respect to $T$ (for brevity we indicate only the derivative of one of
the propagators, the same diagrams with derivatives of another propagator
are included as well).

Fig. 4. Phase diagram as obtained from two- and many-patch RG analyses for $%
U=2t,$ $t^{\prime }=0,$ and $n=1.$ Bold lines bound the weak-coupling region
of the phase diagram, where the RG approach is applicable. Solid lines
correspond to the phase boundaries obtained from the two-patch
analysis. CF and SF denote the charge- and spin-flux phases, respectively,
PS is the interaction-induced phase separated state. F, PI, and A denote the
possibilities for ferromagnetism (F), spontaneous spin-independent (PI) and
spin-dependent (A) deformations of the FS in the corresponding strong
coupling regions. The other notations are the same as in Fig.1. The shaded
region is the frustrated area with the critical scaling parameter of the
two-patch approach $\lambda _c>20.$ The captions at the phase boundary lines
denote the symmetry of the Hamiltonian (\ref{HVH}) on these lines with
respect to the rotation operators $R_\eta ,$ $R_\pi ,$ $R_\tau $ or $R_A$
and the corresponding ordered states which become equivalent. The symbols
correspond to the results of the many-patch RG approach: circles correspond
to SDW, diamonds to dSC, squares to SF, triangles to F, crosses to PS, and
stars to CDW phase. The open symbols denote the frustrated regime in
many-patch calculations when no divergence of the coupling constants was
obtained for $\lambda _T=\ln (t/T)/2<4.$ The type of the corresponding
ordering tendency in this case was determined by the largest susceptibility
at $\lambda _T=4.$

Fig. 5. Phase diagram as obtained from the RG calculations for $U=2t,$
$t^{\prime }/t=0.1,$ and $n=0.92.$  Solid and dashed lines correspond to
the phase boundaries obtained from the two-patch analysis and separate
the phases with respect to the scaling behavior of the coupling constants
(see text). The half-filled circle corresponds a to charge-flux
instability in the many-patch RG analysis. Thin crosses bound the bottom
of the region where the charge-flux fluctuations become substantial,
$\chi _{\text{CF}}>(2/3)\chi _{\text{SDW}}$. Other notations are
the same as in Fig. 4.

Fig. 6. Phase diagram as obtained from the RG calculations for $U=2t,$ $%
t^{\prime }/t=0.3,$ and $n=0.72.$ Thin crosses bound the region where charge-%
flux fluctuations become substantial, $\chi _{\text{CF}}>(2/3) \text{max}(
\chi _{\text{SDW}},\chi _{\text{dSC}})$. Other notations are the same as in
Figs. 4,5.

Fig. 7. Scale dependence of the leading order-parameter inverse
susceptibilities in the two-patch RG approach for $U=2t,$ $t^{\prime }/t=0.1,
$ and $n=0.92$: a) $V=J=0;$ b) $V=0,$ $J=0.2U;$ c) $V=0.5U,$ $J=0.$ The
vertical dot-dashed lines mark the scaling parameter $\lambda ^{*}$ where
the largest absolute value of the coupling constants $|g_i|$ is unity.

\newpage

\psfig{file=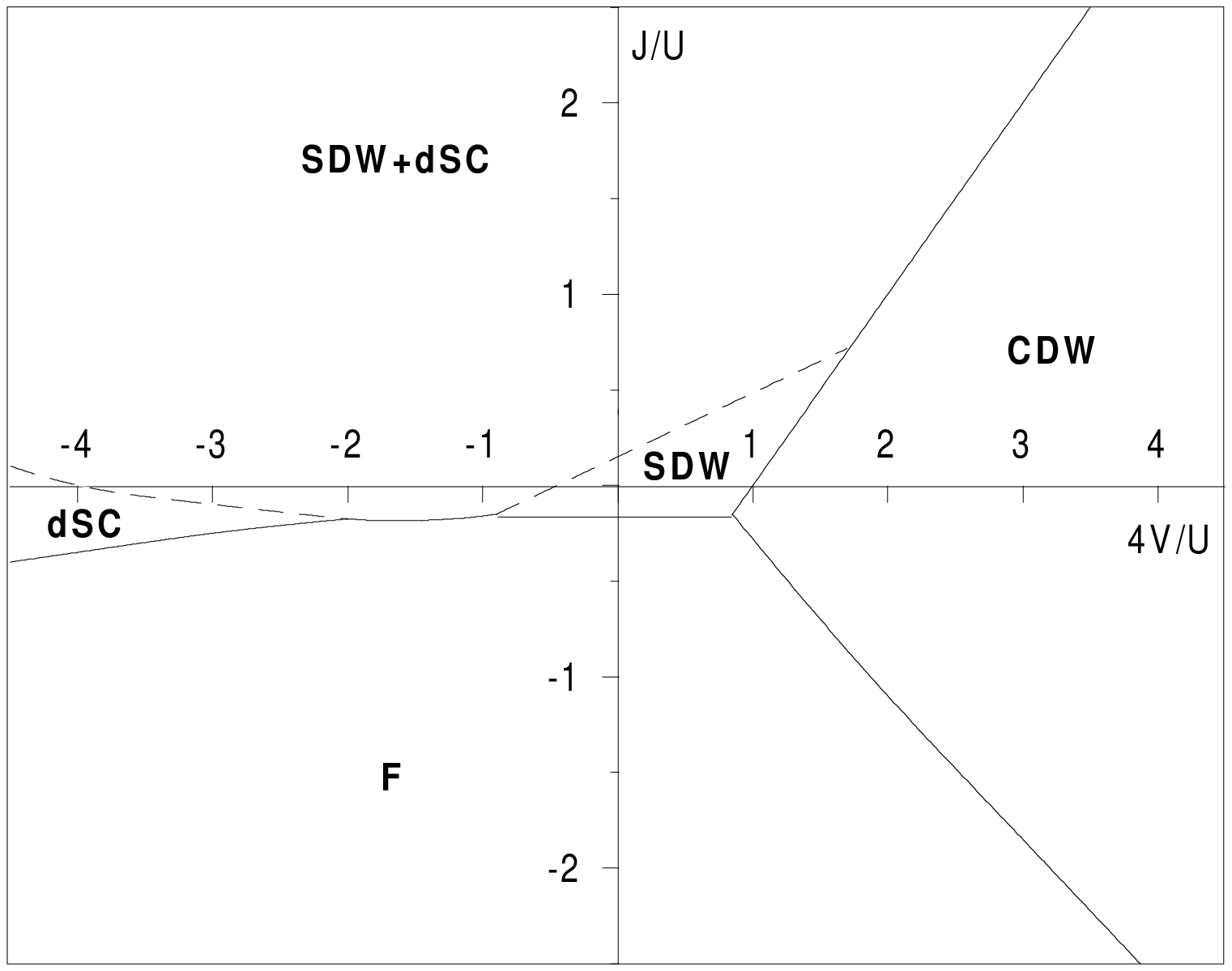}

\newpage

\psfig{file=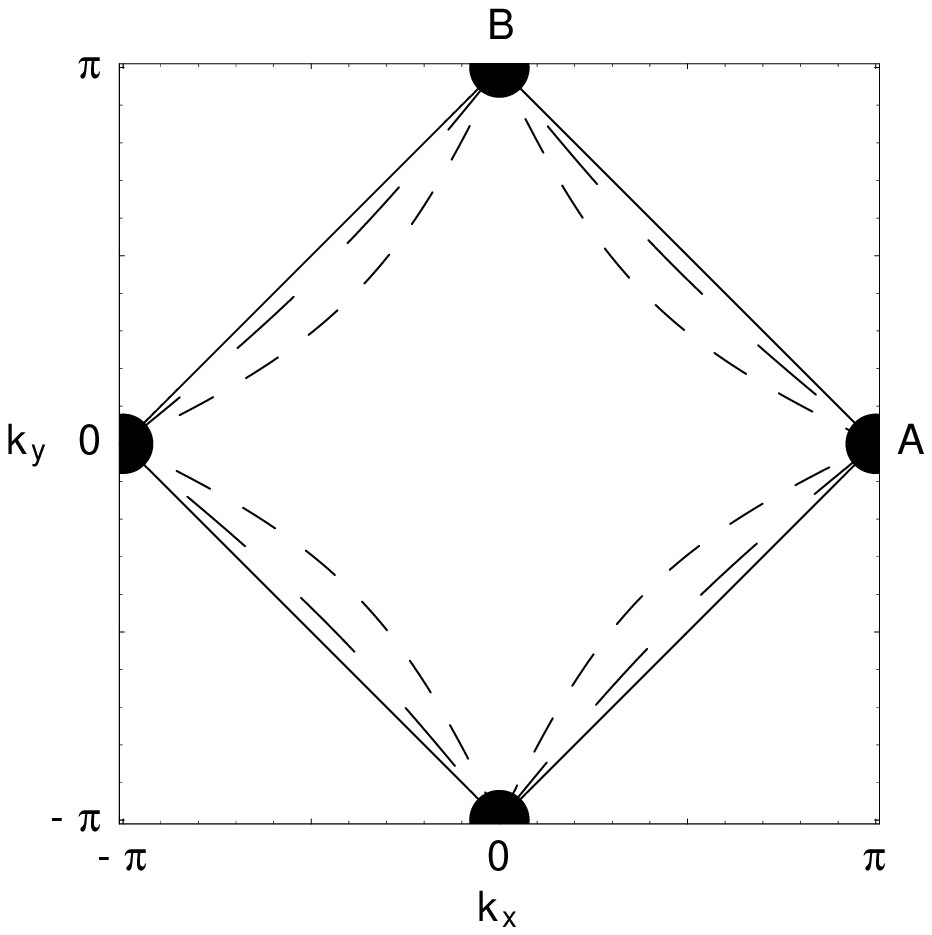}

\newpage

\psfig{file=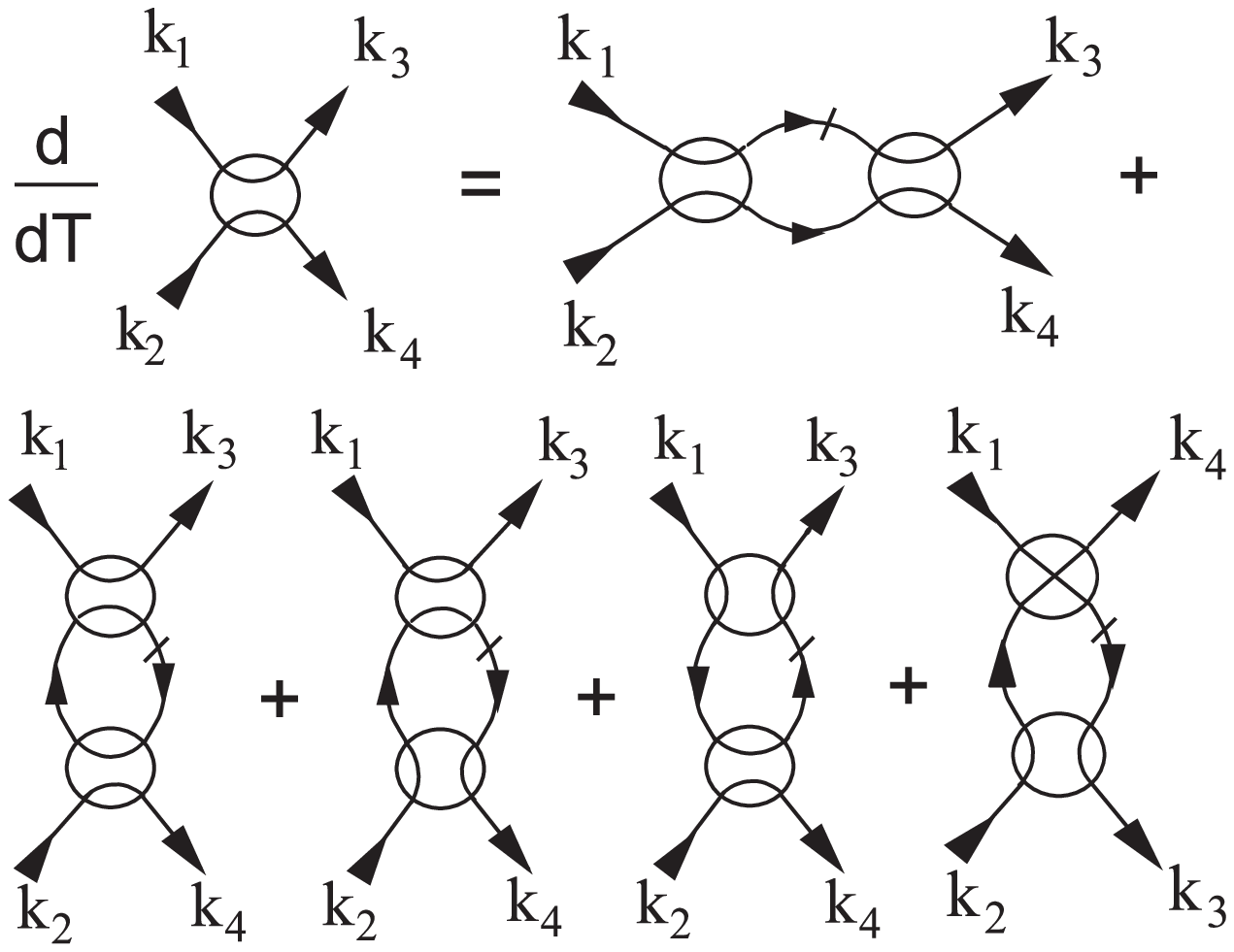}

\newpage

\psfig{file=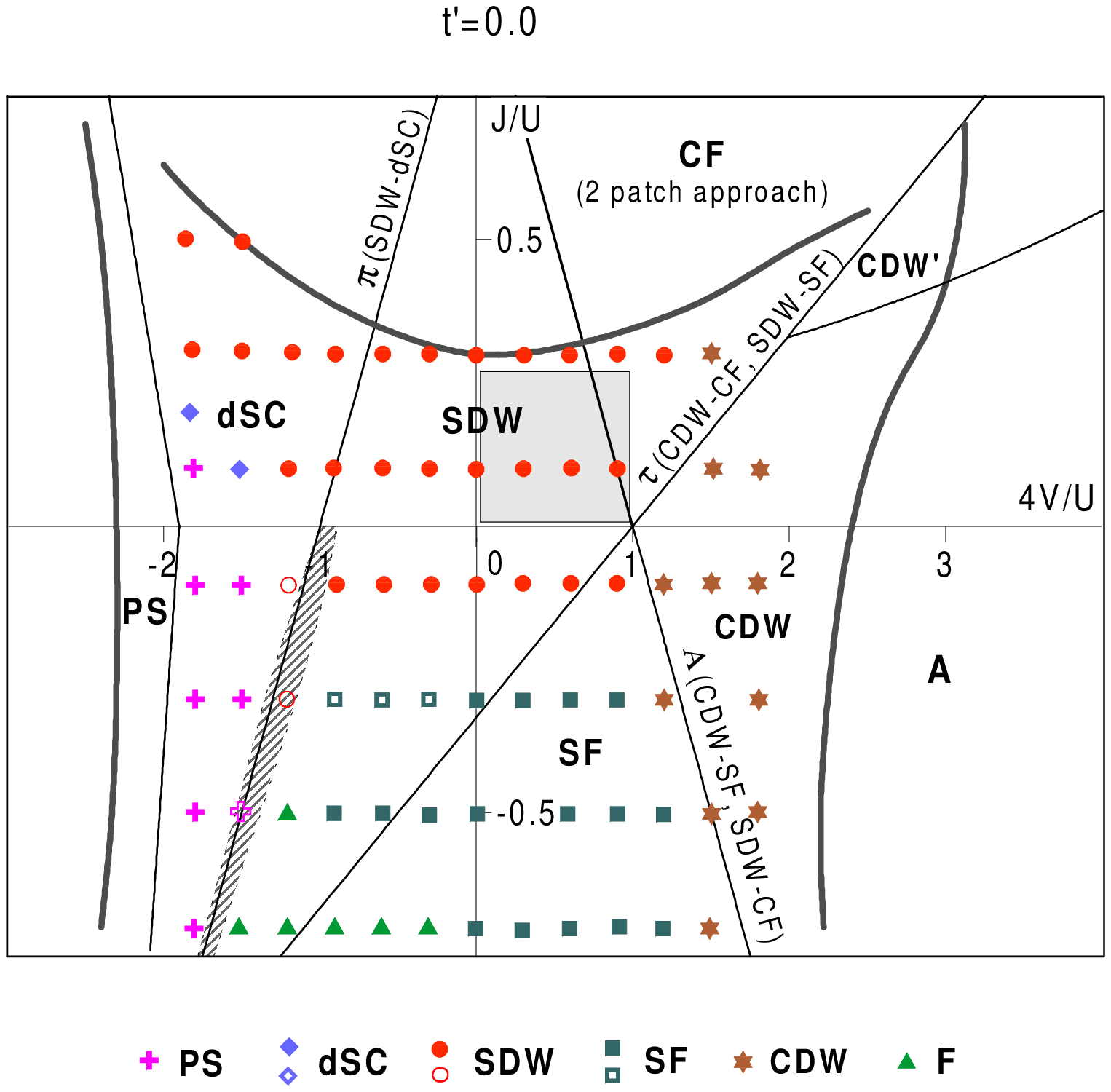}

\newpage

\psfig{file=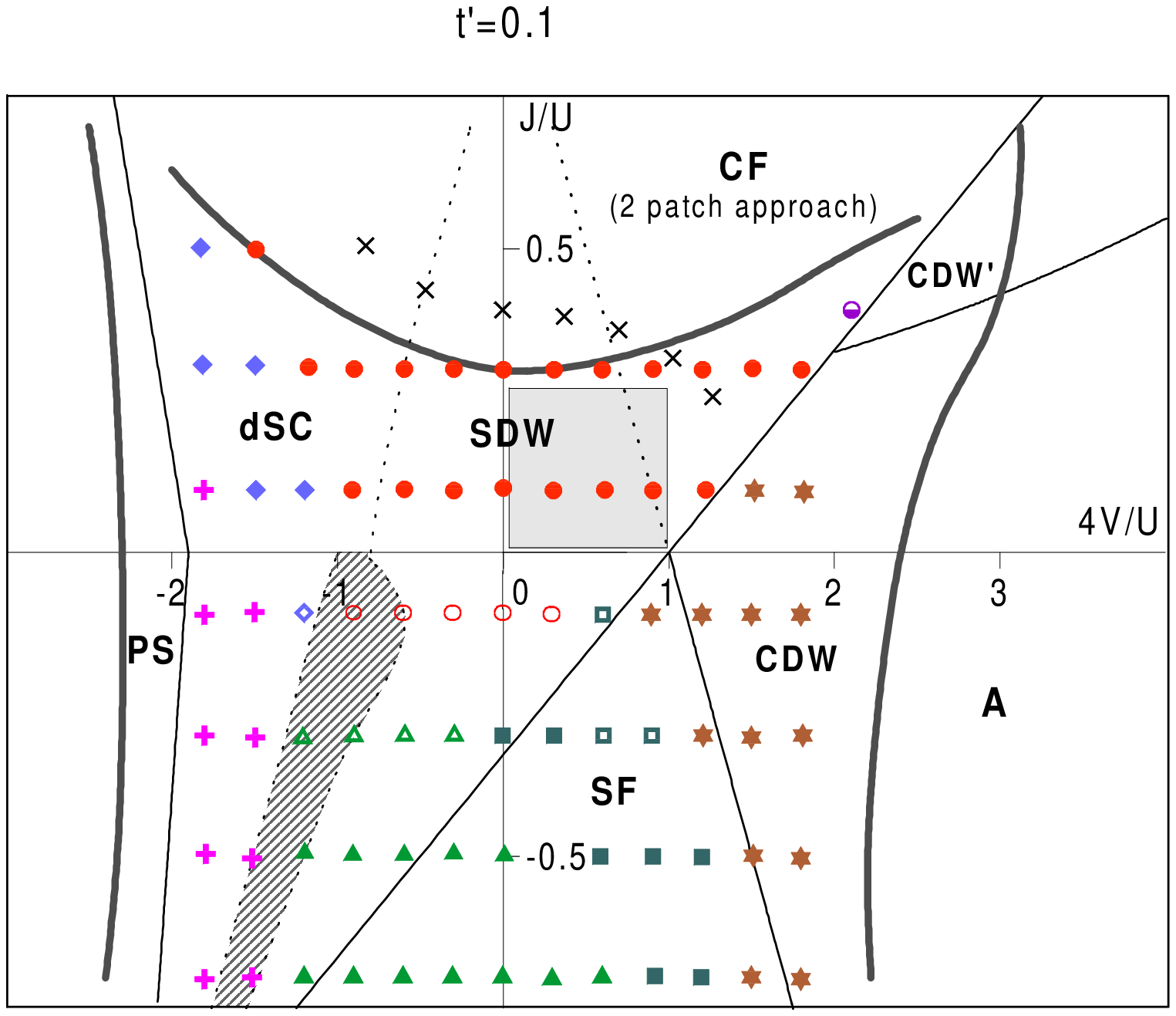}

\newpage

\psfig{file=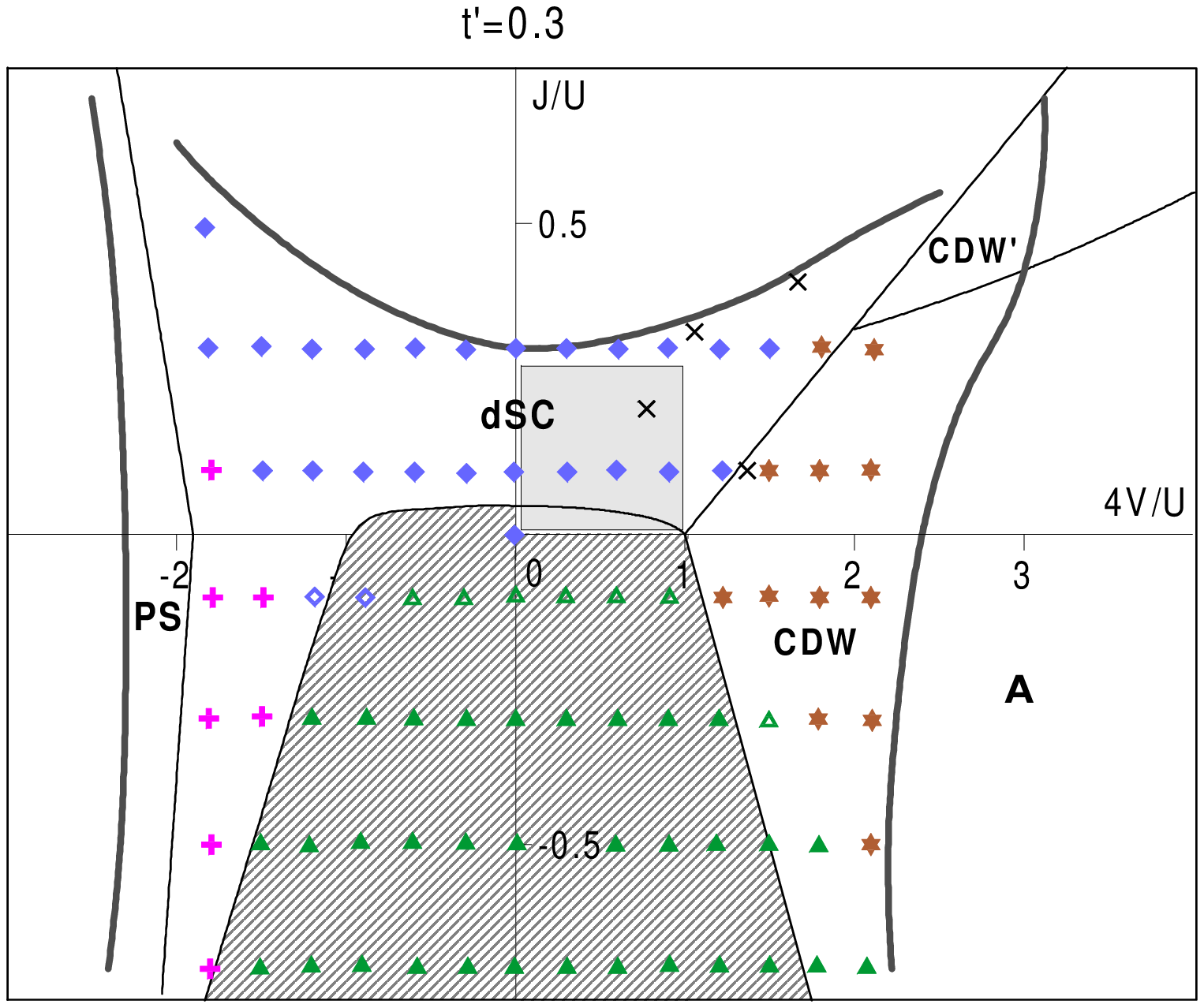}

\newpage

\psfig{file=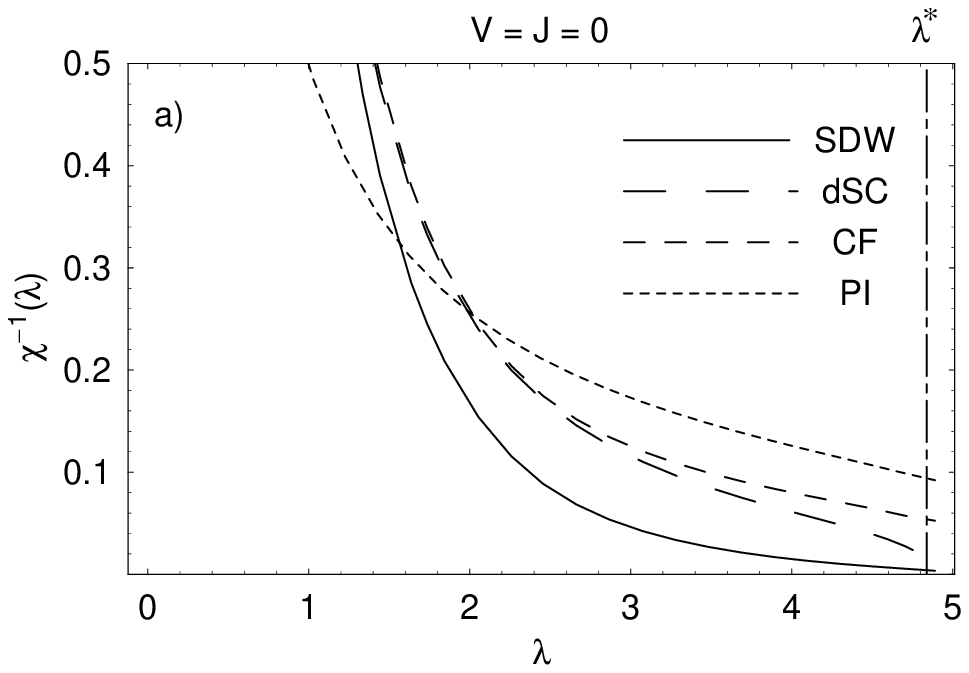}
\vspace{0.1in}
\psfig{file=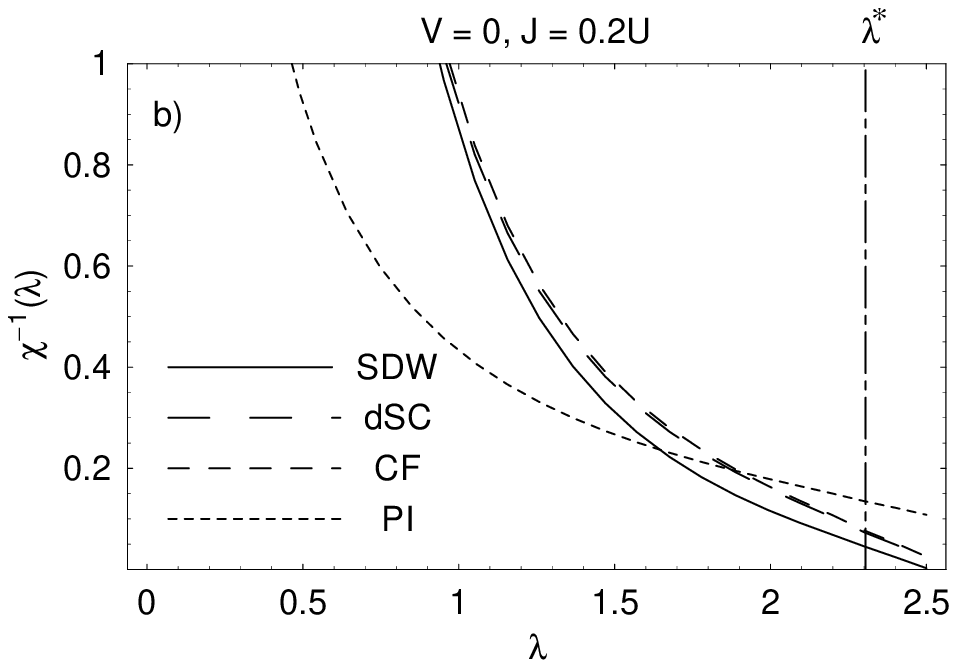}
\vspace{0.1in}
\psfig{file=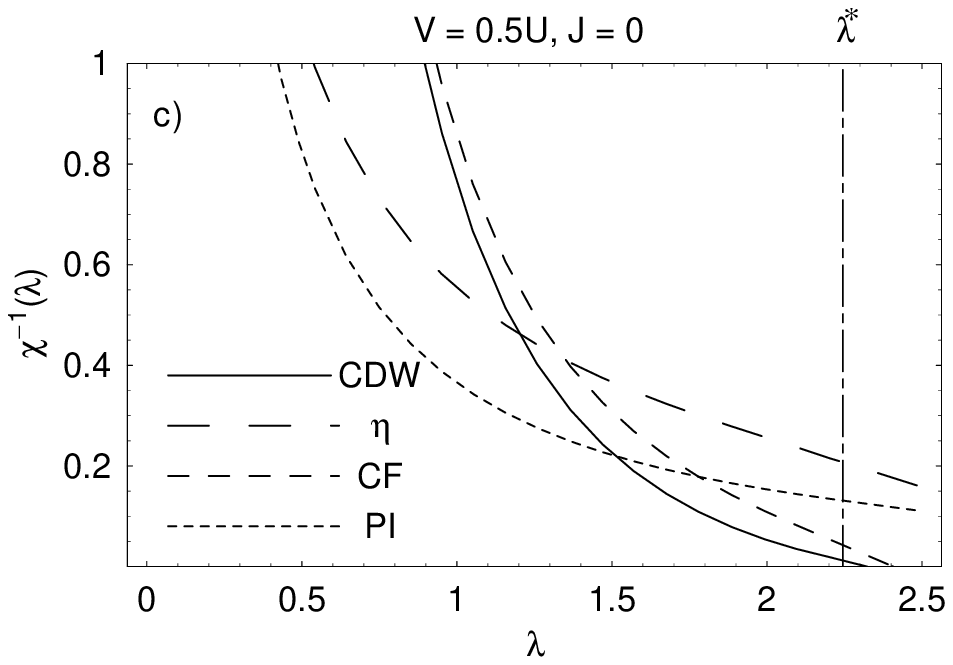}

\end{document}